\documentclass[aip, jmp, amsmath, amssymb, preprint]{revtex4-1}

\usepackage[utf8]{inputenc}
\usepackage[T1]{fontenc}
\usepackage{lmodern, textcomp}
\usepackage[english]{babel}
\usepackage{csquotes}

\usepackage{makecell}

\usepackage{amsmath}
\usepackage{amsfonts}
\usepackage{amssymb}

\usepackage[pdftex, breaklinks=true, linktocpage,
	colorlinks=true, urlcolor=blue, linkcolor=blue, citecolor=red
]{hyperref}

\usepackage{stmaryrd}

\setcellgapes{1pt}
\makegapedcells
\newcolumntype{R}[1]{>{\raggedleft\arraybackslash }b{#1}}
\newcolumntype{L}[1]{>{\raggedright\arraybackslash }b{#1}}
\newcolumntype{C}[1]{>{\centering\arraybackslash }b{#1}}
\newcommand{\Tr}{\operatorname{Tr}}
\newcommand{\tr}{\operatorname{tr}}
\newtheorem{theorem}{Theorem}
\newtheorem{definition}{Definition}
\newtheorem{proposition}{Proposition}
\newtheorem{remark}{Remark}

\newcommand{\droit}{\mathrm{d}}
\newcommand{\p}{\mathrm{p}}

\newtheorem{lemma}{Lemma}

\usepackage{enumerate}
\usepackage{graphicx}

\newcommand{\beq}{\begin{equation}}
\newcommand{\eeq}{\end{equation}}
\newcommand{\bea}{\begin{eqnarray}}
\newcommand{\eea}{\end{eqnarray}}

\newcommand{\bes}{\begin{eqnarray}}
\newcommand{\ees}{\end{eqnarray}}

\newcommand\restr[2]{{
  \left.\kern-\nulldelimiterspace 
  #1 
  \vphantom{\big|} 
  \right|_{#2} 
  }}

\newcommand{\SO}{\mathrm{SO}}

\numberwithin{equation}{section}

\hypersetup{
	pdftitle={Constructive expansion for vector field theories. I. Quartic models in low dimensions}
}

\begin{document}

\title{Constructive expansion for vector field theories \\ I. Quartic models in low dimensions}

\author{Harold Erbin}\email{harold.erbin@gmail.com}
\affiliation{Arnold Sommerfeld Center for Theoretical Physics, Ludwig--Maximilians--Universität, Theresienstraße 37, 80333 München, Germany}

\author{Vincent Lahoche}\email{vincent.lahoche@cea.fr}
\affiliation{\textsc{Cea}/\textsc{List}/\textsc{Diasi}/\textsc{Lvic}, F-91191 Gif-sur-Yvette, France}

\author{Mohamed Tamaazousti}\email{mohamed.tamaazousti@cea.fr}
\affiliation{\textsc{Cea}/\textsc{List}/\textsc{Diasi}/\textsc{Lvic}, F-91191 Gif-sur-Yvette, France}

\date{\today}

\begin{abstract}
This paper is the first of a series aiming to use the loop vertex expansion (LVE) to recover or prove analyticity and Borel summability for generic vector models with bosonic or fermionic statistics in various dimensions. We consider both non-relativistic and relativistic bosons and fermions coupled with a constant quartic tensor in zero-, one- and two-dimensional space,  by limiting our investigations to the super-renormalizable models. This offers a unified perspective on classical constructive results, highlighting the usefulness of the LVE as a modern tool to address these questions and to tackle more challenging models in higher dimensions. Finally, we investigate the large $N$ and massless limits along with quenching for fermions in one dimension. In particular, this work establishes the Borel summability of the Sachdev-Ye-Kitaev model.
\end{abstract}

\maketitle

\section{Introduction}
\label{sec1}

Constructive quantum field theory (CQFT) is a branch of mathematical physics that aims to investigate whether quantum or statistical field theory models are well-defined in a rigorous mathematical sense. Despite their experimental successes, quantum field theories (QFTs) may be badly defined mathematically. This defect was pointed out since their origin almost a century ago. Historically, a vast amount of QFT axioms have been designed since 1950~\cite{Bardakci:1961vt,summers2012perspective}, the most popular and minimal being the \textit{Wightman axioms}~\cite{Streater:2000:PCTSpinStatistics, Flato:1972hu, Glimm:1972zz}. The proof of Wightman axioms for QFT models is reputed to be a difficult task, the difficulty increasing generally with the realism of the considered field theory. All these pioneer works took place in the operator formalism. The development of Feynman's functional integral provided an important change at the beginning of 1970~\cite{Glaser:1974hy, Frohlich:1974ejk}. In Euclidean time, the connection between Wightman and Schwinger functions, a consequence (for instance) of the Osterwalder-Schrader (OS) analytic continuation procedure~\cite{Glaser:1974hy, Osterwalder:1973dx, Benfatto:2006ey, Glimm:1979zi, Glimm:1977xt}, leads to the well-known OS axiomatization. These axioms provide a set of minimal conditions (OS positivity, clustering, ergodicity, etc.) ensuring the possibility to analytically continue from Euclidean Schwinger functions to Wightman functions satisfying all the Wightman conditions. Nowadays, a large literature exists on this subject, especially on the connections between the Euclidean and Minkowskian theories.

At this stage of history, a new type of question emerged in CQFT: “Can we obtain a rigorous definition of quantities like Schwinger functions for interacting models?” From a perturbative point of view, the expansion usually fails to be well-defined, and even for models without divergences or renormalizable~\cite{Rivasseau:1991:PerturbativeConstructiveRenormalization}. For instance, the free energy $F$ may be formally computed in terms of amplitudes $A_{\mathcal G}$ indexed by Feynman diagrams $\mathcal G$:
\begin{equation}
\label{defpert}
F = \sum_{\mathcal{G}\,\text{connected}} A_\mathcal{G}\,.
\end{equation}
However, in general, this sum fails to converge $\sum_\mathcal{G} \vert A_\mathcal{G}\vert =\infty$ because the number of connected Feynman diagrams grows very fast with the number of vertices~\cite{Dyson:1952:DivergencePerturbationTheory} -- a bad consequence coming from the illegal permutation between sums and integrals in the perturbative expansion. The central success of constructive theory on this question was to realize that, using sophisticated methods such as Mayer expansion and lattice clustering~\cite{Brydges:1987:MayerExpansionsHamiltonJacobi, Sokal:1980:ImprovementWatsonsTheorem}, the existence of a non-empty radius of convergence and Borel summability of the perturbative series can generally be proven. The loop vertex expansion (LVE) is a recent constructive technique~\cite{Rivasseau:2007:ConstructiveMatrixTheory, Magnen:2008:Constructivephi4Field, Gurau:2008:TreeQuantumField, Rivasseau:2014:HowResumFeynman, Gurau:2014:RenormalizationAdvancedOverview}: starting with a Hubbard-Stratonovich transformation~\cite{Stratonovich:1957:MethodCalculatingQuantum, Hubbard:1959:CalculationPartitionFunctions}, it replaces the perturbative expansion in terms of Feynman graphs by an expansion indexed by trees, whose number increases very slowly with the number of vertices in comparison~\cite{Rivasseau:2014:HowResumFeynman}. The strategy may be summarized as follows: for any pair made out of a connected Feynman diagram $\mathcal G$ and a spanning tree $\mathcal T \subset \mathcal{G}$, there exists a universal normalized \emph{non-trivial weight} $\varpi(\mathcal{G},\mathcal{T})$ such that:
\begin{equation}
\sum_{\mathcal{T}\subset \mathcal{G}} \varpi(\mathcal{G},\mathcal{T})=1\,,
\end{equation}
which allows to rewrite the original sum \eqref{defpert} as a formal sum over trees
\begin{equation}
\sum_{\mathcal{T}} A_{\mathcal{T}}:=\sum_\mathcal{G} \sum_{\mathcal{T}\subset \mathcal{G}} \varpi(\mathcal{G},\mathcal{T}) A_\mathcal{G},
\end{equation}
and, in the best case, to prove that $\sum_{\mathcal{T}} \vert A_{\mathcal{T}}\vert < \infty$ inside a non-empty domain. In more details, the general strategy for proving Borel summability with LVE uses the Nevanlinna-Sokal theorem, which requires essentially two conditions:
\begin{itemize}
\item the existence of a finite radius of convergence;
\item a bound for the remainder in the Taylor expansion.
\end{itemize}
Since its beginning, the LVE has been successfully applied to a vast amount of problems, in particular, to scalar and matrix models~\cite{Rivasseau:2007:ConstructiveMatrixTheory, Magnen:2008:Constructivephi4Field, Gurau:2008:TreeQuantumField, Rivasseau:2010:LoopVertexExpansion, Lionni:2018:NoteIntermediateField, Rivasseau:2014:HowResumFeynman, Gurau:2014:RenormalizationAdvancedOverview, Gurau:2015:AnalyticityResultsCumulants}, non-commutative field theories~\cite{Wang:2018:ConstructiveRenormalization2dimensional,Rivasseau:2015:CorrectedLoopVertex}, and, more recently, tensor models and tensor field theories, and discrete gravity models~\cite{Gurau:2014:1NExpansionTensor, Delepouve:2016:ConstructiveTensorField, Delepouve:2016:UniversalityBorelSummability, Lahoche:2018:ConstructiveTensorialGroup-1, Lahoche:2018:ConstructiveTensorialGroup-2, Rivasseau:2016:ConstructiveTensorField, Geloun:2018:RenormalizableSYKtypeTensor}.
In particular, a non-trivial improvement over the orthodox LVE, called multi-scale loop vertex expansion (MLVE), has been introduced~\cite{Gurau:2015:MultiscaleLoopVertex, Abdesselam:1995:TreesForestsJungles, Abdesselam:1997:ExplicitFermionicTree} to deal with models having a finite set of divergent graphs (see specifically~\cite{Lahoche:2018:ConstructiveTensorialGroup-2, Delepouve:2016:ConstructiveTensorField, Rivasseau:2019:ConstructiveTensorField}). Finally, the LVE and MLVE do not seem to be the end of the history: a recent paper~\cite{Rivasseau:2017:LoopVertexExpansion} proposed a generalization called loop vertex representation (LVR) which allows to deal with higher-order interactions and which has been successfully applied to random matrix theories~\cite{Krajewski:2017:ConstructiveMatrixTheory, Krajewski:2019:ConstructiveMatrixTheory-1, Krajewski:2019:ConstructiveMatrixTheory-2}.

This paper is the first of a series aiming to prove Borel summability theorems for a large set of Euclidean vector models using specifically the LVE and its generalizations (MLVE, LVR, and so on).
These models and the results we obtain are relevant for machine learning~\cite{Lin:2017:WhyDoesDeep, Bradde:2017:PCAMeetsRG}, condensed matter~\cite{Weinberg:2005:QuantumTheoryFields-2} and SYK models~\cite{Kitaev:2015:SimpleModelQuantum-1, Kitaev:2015:SimpleModelQuantum-2, Maldacena:2016:CommentsSachdevYeKitaevModel, Polchinski:2016:SpectrumSachdevYeKitaevModel} (see~\cite{Rosenhaus:2018:IntroductionSYKModel, Sarosi:2018:AdS2HolographySYK}
for recent reviews), (toy models of) string field theory~\cite{deLacroix:2017:ClosedSuperstringField}, and multi-Higgs models in particle physics phenomenology~\cite{Martin:1997:SupersymmetryPrimer, Binetruy:2007:Supersymmetry, Bento:2017:MultiHiggsDoubletModels}.
For this first work, we focus on the LVE method and general quartic just-renormalizable models with Bose and Fermi statistics. The LVE provides a unified framework to study models having Schrödinger (non-relativistic) and Klein-Gordon / Dirac (relativistic) types of propagators, and arbitrary interactions such that our analysis exhaust all the possible cases of renormalizable models with a purely quartic interaction (up to constraints arising from the stability of the theory). We adopt a presentation where the difficulty is increased step by step with the spatial dimension, the UV divergences becoming more and more difficult to control. For this paper, we focus on models having a finite number of UV divergences: we will study the analyticity and Borel summability of super-renormalizable bosonic and fermionic models in $d = 0, 1$, and relativistic bosonic models in $d = 2$.

Note that this work is not the first one on the subject, the constructive aspects of several bosonic and fermionic models having been investigated in the past. Relativistic bosons, for instance, have been studied in~\cite{Loeffel:1969:PadeApproximantsAnharmonic, Simon:1970:CouplingConstantAnalyticity, Graffi:1970:BorelSummabilityApplication} for $d = 0$; in~\cite{Billionnet:1993:AnalyticityBorelSummability} for $d = 1$, in~\cite{Simon:1970:BorelSummabilityGroundState, Eckmann:1974:DecayPropertiesBorel, Eckmann:1979:BorelSummabilityMass, Rivasseau:2015:CorrectedLoopVertex} for $d = 2$, and in~\cite{Magnen:1977:PhaseSpaceCell,deCalan:1981:LocalExistenceBorel, Magnen:2008:Constructivephi4Field} for $d=3,4$. A general but less rigorous argument for all super-renormalizable relativistic scalar fields with generic polynomial interactions has been given in~\cite{Serone:2017:PowerPerturbationTheory, Serone:2018:lambdaPhi4Theory}.
The Borel summability in the large $N$ limit is discussed in~\cite{Billionnet:1982:AnalyticInterpolationBorel, Frohlich:1982:BorelSummability1N}, and the non-relativistic case in~\cite{Ginibre:1980:ClassicalFieldLimit}. Fermions at finite temperature were investigated for instance in~\cite{Gawedzki:1985ic, Gawedzki:1985ez, Gawedzki:1985uq,Benfatto:2004ib,Lesniewski:1987ha}. Other standard references and textbooks on this subject are~\cite{Mastropietro:2008zz, Salmhofer:1999uq,Rivasseau:1991:PerturbativeConstructiveRenormalization,Glimm:1977xt}
The main addition of our paper is to offer a completely unified perspective using the modern method of LVE.

In this paper, we prove the existence of a constructive expansion for general super-renormalizable fermionic and bosonic quartic models in low dimensions. Then, we extend the results to the quenched Sachdev-Ye-Kitaev (SYK) model~\cite{Kitaev:2015:SimpleModelQuantum-1, Kitaev:2015:SimpleModelQuantum-2, Maldacena:2016:CommentsSachdevYeKitaevModel, Polchinski:2016:SpectrumSachdevYeKitaevModel}.
This is the main new result of this paper since the existence of a constructive expansion for SYK models has not been addressed previously.
While the cases of fermionic and bosonic fields in low dimensions have been largely investigated in the past, we stress again that our paper is the first to deal with them using the LVE and to provide a unified perspective with general quartic interactions.

As explained, this paper is the first of a series, and some questions remain open. Among them, the case of higher-order interactions, which we expect to be treated using improved versions of the LVR (which do not exist today); and the case of just renormalizable interactions. The last case is a difficult task and is also an open problem in tensorial field theories~\cite{Lahoche:2018:ConstructiveTensorialGroup-2, Delepouve:2016:ConstructiveTensorField, Rivasseau:2019:ConstructiveTensorField} where the LVE has been used successfully.\footnote{%
Because the $\phi^4$ model in $d = 4$ is not asymptotically free, the theory is not expected to be Borel summable when integrating along with the flow (Landau pole). This reflects the fact that the theory does not exist by itself, but it can be coupled with other fields. For this reason, the structure of the interaction must be slightly modified to make the beta function negative, which can be understood as writing an effective theory. See~\cite{Rivasseau:2015:WhyAreTensor} for a comparison of the vector, matrix, and tensor cases.
}

The details on the models that we consider and the main statements are given in Section \ref{section2}, the proofs being relegated in Section \ref{section3}.
Appendix \ref{appB} describes the BKAR forest formula, and Appendix \ref{appC} Nevanlinna theorem.

\section{The models, basics and main statements} \label{section2}

In this section, we define the class of models considered in this paper and fix the notations. For convenience, the proofs are given in Section \ref{section3}. We recall some useful properties of $d$-dimensional bosonic and fermionic vector models, including in particular a description of the propagators and the respective power-counting theorems. Finally, we outline the main statements of this paper.

\subsection{The models and power counting}

This paper is devoted to a family of statistical models for random $N$-component vector fields $\{\chi_{i}\}$ in $d$ dimensions, $\chi_{i}:\mathbb{R}^d\to \mathbb{R},\mathbb{C}$, $i=1,\cdots,N$, collectively denoted as $\Xi:=\{\chi_{i}\}$.\footnotemark{}
\footnotetext{%
The index $i$ is sometimes called "flavor".
}%
The value at the point $\textbf{x}\in \mathbb{R}^d$ is denoted as $\chi_i(\textbf{x})$. The field components $\chi_i$ can be either ordinary commutative numbers ($\chi_i\chi_j=\chi_j\chi_i$) or non-commutative (Grassmann valued) numbers ($\chi_i\chi_j=-\chi_j\chi_i$), depending if we describe \emph{bosonic} or \emph{fermionic} random degrees of freedom. For the rest of this paper and to clarify the notations, we will denote as $\phi_i$ the bosonic variables and as $\psi_i$ the fermionic ones. $\Phi=\{\phi_i\}$ then denotes the collective bosonic vector, and $\Psi:=\{\psi_i\}$ the analogous fermionic vector.

We focus on statistical models, that is to say, quantum field theory in imaginary (Euclidean) time. Without loss of generality, the probability law $\p(\Xi)$ for a configuration $\Xi$ is fixed by the choice of a \emph{classical action} $\mathcal{S}(\Xi)$:
\begin{equation}
\p(\Xi)= \frac{1}{Z} \, e^{-\mathcal{S}(\Xi)}\,,\label{def1}
\end{equation}
the \emph{partition function} $Z$ ensuring the normalization of the probability distribution,
\begin{equation}
Z:=\int \left[\prod_{i=1}^N \droit\chi_{i}\right] \, e^{-\mathcal{S}(\Xi)}\,,
\end{equation}
where $\droit\chi_{i}$ denotes the generalized Lebesgue measure.

In this paper, we limit our investigations to quartic models, the first non-Gaussian contributions compatible with $\mathbb{Z}_2$-symmetry. For real fields $\chi_{i}:\mathbb{R}^d\to \mathbb{R}$, we choose a classical action of the form:
\begin{equation}
\mathcal{S}(\Xi)=\int\droit\textbf{x} \left[\frac{1}{2}\sum_{i,j}\chi_{i}(\textbf{x})C^{-1}_{ij} \chi_{j}(\textbf{x})+\frac{u}{4!} \sum_{ijkl}\mathcal{W}_{ijkl}\chi_{i}(\textbf{x})\chi_{j}(\textbf{x})\chi_{k}(\textbf{x})\chi_{l}(\textbf{x})\right]\,.\label{classicaction}
\end{equation}
In this expression, $C_{ij}$ denotes the covariance matrix, $u$ the coupling constant and $\mathcal{W}_{ijkl}$ the \emph{coupling tensor}. Note that the interaction is \emph{local} in the usual meaning in field theory. Moreover, in these notations, $C_{ij}$ is understood as a differential operator with respect to the space variable $\textbf{x}$:
\begin{equation}
C_{ij}^{-1}:=M_{ij}+a^{(1)}_{ij,\alpha}\frac{\partial}{\partial x_\alpha}+ a^{(2)}_{ij,\alpha\beta}\frac{\partial^2}{\partial x_\alpha\partial x_\beta}+\cdots\,,\label{propgeneral}
\end{equation}
where the Greek indices $\alpha$, $\beta$ run from $1$ to $d$ and the Einstein convention for repeated indices is assumed. Note that the \emph{mass matrix} $M_{ij}$ as well as the weight matrices $a^{(n)}_{ij}$ do not depend on the positions $\textbf{x}$. For bosonic fields, $\chi_i\chi_j$ and $\chi_i\chi_j\chi_k\chi_l$ are completely symmetric tensors, whereas they are completely antisymmetric tensors for fermionic fields such that:
\begin{itemize}
\item \emph{bosonic} fields: $C_{ij}=C_{ji}$ and $\mathcal{W}_{ijkl}$ is completely symmetric.

\item \emph{fermionic} fields: $C_{ij}=-C_{ji}$ and $\mathcal{W}_{ijkl}$ is completely anti-symmetric.
\end{itemize}

For complex fields $\chi_{i}:\mathbb{R}^d\to \mathbb{C}$, we choose the classical action to be of the form:
\begin{equation}
\mathcal{S}(\Xi)=\int \droit\textbf{x} \left[\sum_{i,j}\bar{\chi}_{i}(\textbf{x})C^{-1}_{ij} \chi_{j}(\textbf{x})+\frac{u}{4!} \sum_{ijkl}\mathcal{W}_{ijkl}\chi_{i}(\textbf{x})\bar{\chi}_{j}(\textbf{x})\chi_{k}(\textbf{x})\bar{\chi}_{l}(\textbf{x})\right]\,.\label{classicaction-complex}
\end{equation}
Once again, the permutation symmetries of the covariance and coupling tensors depend on the nature of the fields. Requiring the classical action to be a real functional of the fields, we get the conditions:
\begin{itemize}
\item \emph{bosonic} fields: $C_{ij}=C^{\dagger}_{ij}$ and:
\begin{equation}
\mathcal{W}_{ijkl}=\mathcal{W}_{klij}=\mathcal{W}_{ilkj}=\mathcal{W}_{kjil} =\mathcal{W}_{jilk}^*\,.
\end{equation}
\item \emph{fermionic} fields: $C_{ij}=-C^{\dagger}_{ij}$ and:
\begin{equation}
\mathcal{W}_{ijkl}=\mathcal{W}_{klij}=-\mathcal{W}_{ilkj}=-\mathcal{W}_{kjil} =\mathcal{W}_{jilk}^*\,.
\end{equation}
\end{itemize}
Without much loss of generalities, we assume that the propagator takes the form of a direct product between a purely spatial part and a purely internal flavor space:
\begin{equation}
C_{ij}^{-1}:=K_{ij}^{-1}\tilde{C}^{-1}\,,
\end{equation}
choosing the spatial part of the propagator $\tilde{C}$ among three kinds of models relevant for physics and compatible with standard axioms of field theory:
\begin{itemize}
\item \emph{Schrödinger} propagator: non-relativistic bosons and fermions in $(d-1)$-dimen\-sional Euclidean space:
\begin{equation}
\tilde{C}^{-1}=\frac{\partial}{\partial t}-\kappa^* \Delta +m
\end{equation}
where $\Delta$ is the usual Laplacian over $\mathbb{R}^{d-1}$: $\Delta:=\sum_{i=1}^{d-1} \frac{\partial^2}{\partial x_i^2}$, and $\kappa^*:=\frac{\hbar^2}{2m^{*}}$ is a positive constant involving the effective mass $m^{*}$.

\item \emph{Klein-Gordon} propagator: relativistic bosons in $d$-dimensional Euclidean space-time:
\begin{equation}
\tilde{C}^{-1} = -\frac{\partial^2}{\partial t^2}- \Delta +m^2 \,.
\end{equation}

\item \emph{Dirac} propagator: relativistic fermions in $d$-dimensional Euclidean space-time:
\begin{equation}
\tilde{C}^{-1} = \gamma^\alpha \partial_\alpha+m.
\end{equation}
The Dirac matrices $\gamma^\alpha$ with $\alpha=1,\cdots,d$ satisfy the Dirac algebra $\{\gamma^\alpha,\gamma^\beta\}=2\delta^{\alpha\beta}$ and the condition $\tr(\gamma^{\alpha})=0$ for $d>0$. Dirac matrices are $2^n\times 2^n$ matrices where $d=2n$ (resp. $d = 2 n + 1$) for even (resp. odd) $d$ meaning that relativistic fermions $\psi_i$ have to be understood as $2^n$- anti-commutative vectors: $\psi_i^T=(\psi_{i,1},\cdots,\psi_{i,2^n})$.
\end{itemize}
In all these cases, the mass squared is positive, $m^2 > 0$ (unbroken phase).\footnotemark{}
\footnotetext{%
The method employed in this paper relies heavily on this condition.
Indeed, for $m^2 < 0$, the propagator is not positive definite which does not allow to bound the resolvent as in Lemma \ref{lemmauseful2} or \eqref{boundresolventmlve}.
Very recently, the Borel summability for scalar theories with $m^2 < 0$ have been investigated in~\cite{Serone:2019:lambdaPhi24Theory} by generalizing the methods from~\cite{Serone:2017:PowerPerturbationTheory, Serone:2018:lambdaPhi4Theory}.
}%
The matrix $K_{ij}$, on the other hand, is a real invertible $N\times N$ matrix. Note that, in addition to the external indices $i$ and $j$, the matrix $K$ may depend on some \emph{internal indices} like spin, as it will be the case for \emph{Majorana fermions} or \emph{Dirac fermions}. The choice of this matrix depends on the specificity of the model that we consider, however, we can expect it to be bounded:
\begin{equation}
\vert K_{ij}\vert\leq \kappa\,,\forall i,j\,,\qquad \kappa\in\mathbb{R}\,.
\end{equation}
For large $N$, additional conditions can be necessary to make the sums over loops not too large. With this respect, it may be reasonable to impose a condition such that $K_{ij}$ goes to zero for $i,j \gg 1$. The sums are expected to converge for exponential decay, but it has to be checked for slower decays, especially for power decays:
\begin{equation}
\vert K_{ij}\vert \underset{i,j\gg 1}{\lesssim} \,\min\left(\frac{1}{i^\varepsilon},\,\frac{1}{j^\varepsilon}\right),
\qquad
\epsilon >1.
\end{equation}
In this paper, to simplify the proofs, we will assume that the field components are identically distributed $K_{ij}=\delta_{ij}$, but the conclusions can be easily extended to the generic case $K_{ij} \neq \delta_{ij}$.

We will be interested as well in the case where the coupling tensor has its own degrees of freedom. In particular, we will consider the large $N$ limit of the SYK model, introduced initially in~\cite{Kitaev:2015:SimpleModelQuantum-1, Kitaev:2015:SimpleModelQuantum-2, Maldacena:2016:CommentsSachdevYeKitaevModel, Polchinski:2016:SpectrumSachdevYeKitaevModel} (see~\cite{Rosenhaus:2018:IntroductionSYKModel, Sarosi:2018:AdS2HolographySYK}
for recent reviews). Recently, this model received a lot of attention due to its distinguished features: it is solvable in the strong coupling regime, it displays a near-conformal invariance and it saturates the chaos bound proposed in~\cite{Maldacena:2016:BoundChaos}.
All together, these characteristics make the SYK model a good candidate to describe black holes (with near-horizon geometry $\mathrm{AdS}_2$) through the AdS/CFT correspondence~\cite{Maldacena:1999:LargeNLimit}.
The SYK model, as originally defined in~\cite{Kitaev:2015:SimpleModelQuantum-1, Kitaev:2015:SimpleModelQuantum-2}, corresponds to a $d = 1$ quantum mechanical system (time only) of $N$ Majorana $\psi_i$ with a quartic interaction
\begin{equation}
\mathcal S = \int \droit t \left( \frac{1}{2} \, \psi_i \frac{\droit}{\droit t}\, \psi_i
+ \frac{1}{4!} \, \mathcal J_{ijkl} \, \psi_i \psi_j \psi_k \psi_l
\right).
\end{equation}
where the index $i$ runs from $1$ to $N$, with random couplings (quenching), which is achieved by integrating over $\mathcal J_{ijkl}$ with a Gaussian measure. One considers the quenched partition function:
\begin{equation}
Z(u)=\int \droit\mathcal{J} \,e^{-\mathcal{S}(\mathcal{J})}\int \droit \psi \,e^{-\mathcal{S}(\mathcal{J},\psi)}=:\int \droit\mathcal{J} \,e^{-\mathcal{S}(\mathcal{J})} Z(\mathcal{J})\,,
\end{equation}
where we call $Z(\mathcal{J})$ the \emph{partial partition function}, integrating over the fermionic degrees of freedom only. Once again, note that $\mathcal{J}$ is assumed to be real, with strictly positive eigenvalues $\xi_I^2$. The measure for $\mathcal{J}$, $\droit \mathcal{J} e^{-\mathcal{S}(\mathcal{J})}$ is Gaussian:
\begin{equation}
\mathcal{S}(\mathcal{J}):= \frac{\alpha(N)}{2\bar{g}^2} \sum_{ijkl} \mathcal{J}_{ijkl}\mathcal{J}_{ijkl}\,,
\end{equation}
where $\alpha(N)$ is a certain power of $N$.
Note that the generalized SYK model involves an interaction of degree $q$, $\mathcal{J}_{i_1\cdots i_q} \psi_{i_1} \cdots \psi_{i_q}$: our conclusion about the Borel summability holds only for the original $q = 4$ SYK model.
The function $\alpha(N)$ is fixed to $N^{q-1}$ usually, and we will set $\alpha(N)=N^3$ for our considerations.

The classical action for the fermions is chosen to be:
\begin{equation}
\mathcal{S}(\mathcal{J},\psi):=\int_{-\beta/2}^{\beta/2} \mathcal{L}(\Psi, \mathcal{J})\droit t
\end{equation}
with the Lagrangian density:
\begin{equation}
\mathcal{L}(\Psi, \mathcal{J}):= \frac{1}{2} \sum_i\psi_i \frac{\droit}{\droit t} \psi_i-\frac{u}{4!} \sum_{ijkl} \mathcal{J}_{ijkl} \psi_i\psi_j\psi_k\psi_l\,.\label{sykdef}
\end{equation}
Note the minus sign in front of the interaction term, which arises from $-1 = i^{q/2}$ for $q = 4$, will be crucial for the convergence of the constructive bound.

\medskip

In this paper, we focus on UV finite and super-renormalizable field theories in low dimensions. Power-counting provides a criterion to decide if a theory is UV safe or not (we do not consider IR problems in this paper). The multi-scale analysis is a rigorous method to count the divergences of the theory~\cite{Rivasseau:1991:PerturbativeConstructiveRenormalization}. It introduces a slicing in the interior of the Feynman amplitudes, decomposing each propagator edge into slices. More concretely, regularizing all the propagators using the Schwinger parametrization, we introduce a slice index $i \ge 1$ and a thickness parameter $M$, in such a way that the full propagator $\tilde{C}$ in Fourier space is given by the sum:
\begin{equation}
\tilde{C}=\sum_{i=1}^{i_{max} }\tilde{C}^{(i)}+\tilde{C}^{(0)}\,,
\end{equation}
the integer $i_{max}$ being fixed.
The sliced propagator $\tilde{C}^{(i)}$ for $i\ge 1$ is given by:
\begin{equation}
\tilde{C}^{(i)}(\omega,\vec{p}\,^2):=(-i\omega+{E}_c)\int_{M^{-2i}}^{M^{-2(i-1)}} \droit\alpha \, e^{-\alpha(\omega^2+{E}_c^2)} \,,\label{nonrel}
\end{equation}
for non-relativistic bosons and fermions, the kinetic energy being $E_c:=\kappa \vec{p}\,^2+m$;
\begin{equation}
\tilde{C}^{(i)}(\omega^2,\vec{p}\,^2):=\int_{M^{-2i}}^{M^{-2(i-1)}} \droit\alpha \, e^{-\alpha(\omega^2+\vec{p}\,^2+m^2)}\,,\label{relbos}
\end{equation}
for relativistic bosons, and
\begin{equation}
\tilde{C}^{(i)}(\omega,\vec{p}\,):= (-\gamma^\mu \partial_\mu+m)\int_{M^{-2i}}^{M^{-2(i-1)}} \droit\alpha \, e^{-\alpha(\omega^2+\vec{p}\,^2+m^2)}\,,\label{relferm}
\end{equation}
for relativistic fermions.\\
In the amplitude decomposition, $M^{i_{max}}$ has to be interpreted as the UV cut-off, ensuring the finiteness of the loop integrals for large momenta. The remainder $\tilde{C}_{ij}^{(0)}$ includes integration over Schwinger parameter $\alpha$ from $1$ to $\infty$, and is completely regular in the UV. From the slice decomposition of the amplitude, it is not hard to show the following statement:
\begin{theorem} \label{theorem1}
Let $\mathcal{A}_{\mathcal{G}}$ be the amplitude of the connected Feynman graph $\mathcal{G}$ with $V$ vertices and $L$ internal propagator edges. Introducing a slicing for each propagator edge, we denote by $\mu=\{i_1,\cdots,i_L\}$ a scale assignment for edges labelled from $1$ to $L$. The amplitude can then be decomposed as a sum over scale assignments such that
\begin{equation}
\mathcal{A}_{\mathcal{G}}=\sum_{\mu} \mathcal{A}_{\mathcal{G},\,\mu}\,.
\end{equation}
For fixed $\mu$, let $\mathcal{G}_i = \bigcup_{k=1}^{k(i)}\mathcal{G}_i^k\subset \mathcal{G}$ be the subgraph of $\mathcal{G}$ made of edges with scale assignments higher or equal to $i$, and $\mathcal{G}_i^k$ its $k$th connected component, and where $k(i)$ is the total number of connected components. We have the following statement in the UV sector, that is, for scale assignments higher or equal to $1$:
\begin{equation}
\vert \mathcal{A}_{\mathcal{G},\,\mu}\vert \leq K^{L(\mathcal{G})} \prod_{i,k} M^{\Omega(\mathcal{G}_i^k)}\,,
\end{equation}
the bound being uniform for some constant $K$.
For a quartic model we have:
\begin{itemize}
\item non-relativistic bosons and fermions:
\begin{equation}
\Omega(\mathcal{G}_i^k)=\frac{1}{2}(d-3)V(\mathcal{G}_i^k)-\frac{d-1}{4}N(\mathcal{G}_i^k)+ \frac{1}{2}(d+1) \label{crit1}\,,
\end{equation}

\item relativistic bosons:
\begin{equation}
\Omega(\mathcal{G}_i^k)=(d-4)V(\mathcal{G}_i^k)-N(\mathcal{G}_i^k)\left(\frac{d}{2}-1\right) +d\label{crit2}\,,
\end{equation}

\item relativistic fermions:
\begin{equation}
\Omega(\mathcal{G}_i^k)=(d-2)V(\mathcal{G}_i^k)-\frac{d-1}{2}N(\mathcal{G}_i^k)+d \label{crit3}\,,
\end{equation}
\end{itemize}
In these expressions, $N(\mathcal{G})$ and $V(\mathcal{G})$ denote respectively the numbers of external edges and vertices of the diagram $\mathcal{G}$.
\end{theorem}
The proof of this theorem is classic, and the reader may consult~\cite{Rivasseau:1991:PerturbativeConstructiveRenormalization} for details. We have to specify the ranges of the momenta $\omega$ and $\vec{p}$, which may be different. For non-relativistic models at finite temperature $\beta$, the time coordinate is periodic $t \sim t + \beta$ and a fundamental interval is $t\in[-\beta/2,\beta/2]$.
Then, the conjugate momentum $\omega$ takes discrete values:
\begin{equation}
\omega=\frac{2\pi n}{\beta}\quad \text{(bosons)},
\qquad
\omega=\frac{(2n+1)\pi}{\beta}\quad \text{(fermions)}\,,\label{frequencies}
\end{equation}
where $n\in\mathbb{N}$.
These two constraints for bosons and fermions come respectively from the usual periodicity and anti-periodicity imposed on Fourier basis functions.
The zero-temperature limit $\beta \to \infty$ corresponds to non-compact Euclidean time $t \in \mathbb{R}$ (without periodic conditions).
The spatial momentum $\vec{p}$, however, is not restricted, except if we impose to the system to leave into a box of finite size. For our purpose, we consider $\vec{p}\in\mathbb{R}^{d-1}$.

We deduce the following renormalizability criteria for each case studied in this paper:
\begin{itemize}
\item \emph{Non-relativistic bosons and fermions} are super-renormalizable for $d<3$, just-renormalizable for $d=3$ and non-renormalizable otherwise.

\item \emph{Relativistic bosons} are super-renormalizable for $d <4$, just renormalizable for $d=4$ and non-renormalizable otherwise.

\item \emph{Relativistic fermions} are super-renormalizable for $d<2$, just renormalizable for $d=2$, and non-renormalizable otherwise.
\end{itemize}

\subsection{Main statements}

We now outline our main results. We call \textit{tree expansion} the expansion of the free energy obtained from BKAR forest formula (Appendix \ref{appB}). We have:

\begin{enumerate}
\item First, for fixed and positive (see definition \ref{defpos}) coupling tensor $\mathcal{W}$,

\begin{proposition}\label{Stat1}
Let $F(u):=\ln (Z(u))$ be the free energy of a quartic model in $d$ dimensions with generic but fixed tensor coupling and $K_{ij}=\delta_{ij}$. Then, for the relativistic fermionic models in $d\leq 1$, non-relativistic bosonic models in dimensions $d\leq 1$ and the relativistic bosonic models in $d\leq 2$, the tree expansion of $F(u)$ is analytic in the interior of the cardioid domain:
\begin{equation}
\vert u \vert \leq \mathcal{O}\left(1\right) \frac{1}{N\lambda_0^2} \cos^2 \left( \frac{\varphi}{2} \right)\,,
\end{equation}
and corresponds to the Borel sum of the perturbative expansion of the free energy $F(u)$.
\end{proposition}

\item Secondly, for disordered coupling with standard deviation $\langle \mathcal{J}^2 \rangle = \bar{g}/N^3$ and Majorana fermions (SYK model), we have:

\begin{proposition} \label{Stat2}
In the large $N$ limit, the tree expansion of the free energy $F(u)$ of the SYK model, for $u=\rho e^{i\varphi}$ is analytic in the interior of the cardioid domain:
\begin{equation}
\rho \leq \mathcal{O}\left( 1\right) \frac{1}{\bar{g}} \cos^2 \left( \frac{\varphi}{2} \right)\,,
\end{equation}
and corresponds to the Borel sum of the perturbative expansion in $u$.
\end{proposition}
\end{enumerate}
The rest of this paper is devoted to the proof of these statements.

Note that we do not consider the non-relativistic bosonic models in dimension $d=2$, even though it is power counting super-renormalizable. The reason is that we expect such divergences to be unphysical because the non-relativistic approximation is valid only with a cut-off.
In dimensions $d\leq 1$, however, no physical divergence is expected. Indeed, the power counting $\Omega=1-V$ provides a logarithmic divergence for $V=1$, coming from formal sums as $\sum_\omega 1/(i\omega+m)$. These divergences however can be removed by remembering that the path integral construction requires time discretization with finite time step $\epsilon$, providing additional oscillations $e^{i\omega \epsilon}$. These oscillations allows to use the standard trick to compute sums like $\sum_\omega g(i\omega)$ by replacing them by a contour integration into the complex plane. For $\beta=2\pi$, the net result is:
\begin{equation}
\sum_{n\in \mathbb{Z}}\, \frac{1}{i\omega_n+y}\equiv\mp \,2\pi \frac{1}{e^{-2\pi y}\mp 1}\,. \label{equationtadpole}
\end{equation}

\section{Proofs}\label{section3}

In this section, we give the proofs of the two statements \ref{Stat1} and \ref{Stat2} using the LVE. Our general strategy for the proofs is to show that the two requirements of Nevanlinna's theorem (recalled in appendix \ref{appC}) hold~\cite{Sokal:1980ey}. We start with bosons and with dimensions $d \le 1$.
The case $d=2$ is technically more difficult asince it requires an improved version of the LVE, called multi-scale loop vertex expansion (MLVE).

\subsection{Bosons in dimensions \texorpdfstring{$d \leq 1$}{d =< 1}}

\subsubsection{Existence of a finite analyticity domain}

We start by proving that the first requirement of Nevanlinna's theorem holds, that is, the existence of a finite analyticity domain. The zero-dimensional model is simpler and captures the dependence on the internal indices. Since there are no UV divergences, the case $d = 1$ follows easily. \\

Let $Z(u)$ be the partition function of the bosonic model in $d = 0$:
\begin{equation}
Z(u)=\int \droit\mu_C(\phi) \, e^{-\frac{u}{4!} \sum_{ijkl}\mathcal{W}_{ijkl}\phi_{i}\phi_{j}\phi_{k}\phi_{l}}\,,
\end{equation}
where $d\mu_C(\phi)$ is a shorthand notation for the normalized Gaussian measure with covariance $C^{-1}$, $\droit\mu_C(\phi):=\prod_{i=1}^N \droit\phi_i \, e^{-\frac{1}{2}\phi_iC^{-1}_{ij} \phi_j}$.
For $d=0$, $C_{ij}\equiv \delta_{ij}$ for i.i.d variables. The interaction tensor can be viewed as a matrix $\mathcal{W}_{IJ}$, where the big indices $I, J\equiv (ij), (kl)$ run from $1$ to $N(N+1)/2$. As a symmetric matrix with real coefficients, $\mathcal{W}_{IJ}$ can be diagonalized with eigenvalues $\{\Lambda_I\}$:
\begin{equation}
\mathcal{W}_{IJ}=\sum_L\Lambda_LO_{IL}O^{T}_{LJ}\,,
\end{equation}
where $O$ is an orthogonal matrix. Defining the new field $\Psi_L:=\sum_{I=(ij)}O_{IL}\phi_{i}\phi_{j}$, the partition function becomes:
\begin{equation}
Z(u)=\int \droit\mu_C(\phi) \prod_{L}e^{-\frac{u}{4!} \Lambda_L [\Psi_L]^2}\,.\label{decomp1}
\end{equation}
If we suppose $u$ to be positive, a simple way to ensure stability is to assume that the eigenvalues $ \Lambda_L$ are all positive.
We adopt the following definition:
\begin{definition}\label{defpos}
A coupling tensor $\mathcal{W}$ is said to be positive definite if all the eigenvalues $\Lambda_L$ are positive.
\end{definition}

In fact, we will remove this condition formally, venturing into the complex plane $u=\rho e^{i\varphi}$, and show that summability makes sense when $u$ becomes complex. Now, we introduce the matrix-like \emph{intermediate fields} $\sigma^{(L)}$, indexed with a pair of indices $L$, with the normalized Gaussian measure:
\begin{equation}
\droit\nu(\sigma):= \frac{\prod_L e^{-\frac{1}{2}(\sigma^{(L)})^2} \droit\sigma^{(L)}}{\int \prod_L e^{-\frac{1}{2}(\sigma^{(L)})^2} \droit\sigma^{(L)}}\,,
\end{equation}
in order to break the quartic interaction $ [\Psi_L]^2$ as a three-body interaction:
\begin{equation}
Z(u)=\int \droit\mu_C(\phi)d\nu(\sigma) \prod_{L}e^{\sqrt{\frac{{-u}}{12} }\lambda_L \Psi_L\sigma^{(L)}}\,,
\end{equation}
where $\lambda_L^2:=\Lambda_L$. Note that $\lambda_L \Psi_L:=A_L$ is a symmetric matrix in the original little indices $L\equiv (kl)$. Therefore, only the symmetric part of the matrix $\sigma^{(kl)}$ is relevant, and the anti-symmetric part of the intermediate field matrix can be integrated out.\footnote{Indeed, decomposing $\sigma$ in symmetric and anti-symmetric parts as $\sigma=\sigma_{S}+\sigma_{AS}$, the kinetic action for the matrix $\sigma$ writes as $\frac{1}{2}\sum_{ij} (\sigma^{(ij)})^2=\frac{1}{2} \Tr (\sigma_S)^2-\frac{1}{2}\Tr (\sigma_{AS})^2$, so that the anti-symmetric part factorizes exactly in front of the path integral, and is canceled by the normalization.} The integration over the original field variable is Gaussian with effective covariance $C_I^{-1}-\sqrt{\frac{-u}{3}} \, \mathcal{O}_{I}(\sigma)$. Performing this integration leads to the factor:
\begin{equation}
\label{eq:0d-eff-action}
\det\left(1-\sqrt{\frac{{-u}}{3} } \, C \mathcal{O}(\sigma)\right)^{-1/2} =e^{-\frac{1}{2}\Tr\ln\left(1-\sqrt{\frac{{-u}}{3} }C\mathcal{O}(\sigma)\right)}\,,
\end{equation}
with $ \mathcal{O}(\sigma):=\sum_L O_{IL}\lambda_L\sigma^{(L)}$. Then, we expand the partition function in powers of the effective vertex $\mathcal{V}( \mathcal{O}(\sigma)):=\frac{1}{2} \, \Tr\ln\left(1-\sqrt{\frac{{-u}}{3} } \, C \mathcal{O}(\sigma)\right)$. Next, we introduce replicas, replacing the intermediate fields $\sigma^{(L)}$ by $n$ vector fields $\sigma^{(L)}_p$, with $p$ running from $1$ to $n$ and with inverse covariance matrix $\textbf{1}$, with all entries equals to $1$. Denoting as $\droit\nu_\textbf{1}(\sigma)$ the corresponding normalized Gaussian measure, and exchanging sums and integrations, the partition function can be rewritten as:
\begin{equation}
Z(u)=\sum_n\int \droit\nu_\textbf{1}(\sigma) \frac{(-1)^n}{n!} \prod_{p=1}^n \mathcal{V}( \mathcal{O}(\sigma)_p)\,.\label{expansionLVE2}
\end{equation}
From the Brydges-Kennedy-Abdes\-selam-Rivasseau (BKAR) forest formula (see Appendix \ref{appB}), and the elementary properties of Gaussian integration, one arrives at the following expression for the free energy:\footnote{Note that the free energy, which is the logarithm of the partition function, is expanded as a sum of trees rather than forests.}
\begin{align}
\nonumber F(u)=\sum_n \frac{1}{n!}\sum_{\mathcal{T}_n}\prod_{\ell\in\mathcal{T}_n}&\left(\int_0^1 \droit x_{\ell}\right)e^{\frac{1}{2} \sum_{pq} \frac{\partial}{\partial \chi^{(I)}_p}\mathcal{W}_{IJ}X_{pq}(x_\ell)\frac{\partial}{\partial \chi^{(J)}_q}} \\
&\prod_{\ell\in\mathcal{T}_n}\frac{1}{2}\frac{\partial}{\partial \chi^{(I)}_{i(\ell)}} \mathcal{W}_{IJ} \frac{\partial}{\partial\chi^{(J)}_{j(\ell)}}\prod_{p=1}^n \mathcal{V}(\chi_p)\Big\vert_{\chi^{(L)}_p=0}\,,
\end{align}
where we defined $\chi^{(L)}_p:=\lambda_L\sigma^{(L)}_p$, and performed the change of variable $\chi$ by $\chi\to \chi^\prime:=O\chi$. The term with $n=0$ will be treated separately, and we define $\tilde{F}(u)=\sum_{n=1}^{\infty} \frac{1}{n!}F_n(u)$. Computing the derivatives, we get:
\begin{align}
\nonumber F_n(u):=\frac{(-1)^n}{2^{n-1}}&\sum_{\mathcal{T}_n}\prod_{\ell\in\mathcal{T}_n}\left(\int_0^1 \droit x_{\ell}\right) \int d\mu_{\varpi\otimes x}(\chi^{(I)}_p) \prod_{v\in\mathcal{T}_n} (c(v)-1)! \left(\sqrt{-\frac{u}{3}}\right)^{c(v)}\\
&\qquad
\prod_{\ell\in\mathcal{T}_n}\varpi_{I_{t(\ell)} I_{s(\ell)}} \prod_{v\in\mathcal{T}_n} \mathcal{V}_{I_{v,1},\cdots,I_{v,c(v)}}(\chi_p)\Big\vert_{\chi^{(L)}_p=0}\,,
\end{align}
where the $s(\ell)$ (resp.\ $t(\ell)$) are sources (resp. targets) of the edge $\ell$, which can be written as couples $s(\ell) = (v(\ell),i)\,\, 1\leq i \leq c(v)$ (and similarly $t(\ell) = (v(\ell),i)$).
The couple $(v,i)$ with $1\leq i\leq c(v)$ denotes the boundaries of each of the $c(v)$ edges hooked to $v$.
The vertices $\mathcal{V}_{I_{v,1},\cdots,I_{v,c(v)}}$ are defined as:
\begin{equation}
\mathcal{V}_{I_{v,1},\cdots,I_{v,c(v)}}:=\left[\prod_{\ell \vert v\in \partial \ell } \frac{\partial}{\partial\chi^{(J)}_{j(\ell)}}\right]\,\mathcal{V}(\chi_v).
\end{equation}
Explicitly, if we denote as $(i_{v,n},j_{v,n}) \equiv I_{v,n}$ the pair of indices labelled with $I_{v,n}$, we have:
\begin{equation}
\mathcal{V}_{I_{v,1},\cdots,I_{v,c(v)}}=\frac{1}{2}\left[ \prod_{n=1}^{c(v)} \delta_{j_{v,n}i_{v,n+1}} \right]\,[R^{c(v)-1}(\chi_v)]_{i_{v,1}j_{v,c(v)}}\,,
\end{equation}
where we introduced the \emph{resolvent}
\begin{equation}
R(\chi):=\frac{1}{1-\sqrt{\frac{{-u}}{3} }\chi} \,. \label{R}
\end{equation}
The following classical Lemma allows to bound the resolvent, and to extend the domain of convergence in the complex plane:
\begin{lemma}\label{lemmauseful2}
Let $\mathcal{E} \equiv \mathbb{C}^N\otimes L_2(S^1,\mathbb{R})$, where $L_2(S^1,\mathbb{R})$ denotes the space of square summable functions spanned by $\{e^{i n t}\}$, with $n\in \mathbb{Z}$ and $t\in (0,1]$. Let $\{\vert j,n\rangle\}$ be an orthogonal basis on $\mathcal{E}$, for $j$ running from $1$ to $N$. Let $\mathcal{H}$ be an Hermitian operator on $\mathcal{E}$. By definition, $\mathcal{H}$ acts on the basis states $\vert i,n \rangle$ as:
\begin{equation}
\mathcal{H}\vert i,n \rangle = \sum_{j,m} \mathcal{H}_{ji,mn} \vert j,m \rangle\,.
\end{equation}
Let $u=\vert u \vert e^{i\varphi}$, $\varphi\in (-\pi,\pi]$ and $R$ be the resolvent, defined as
\begin{equation}
R^{-1}=\mathrm{id}+i\sqrt{u} \mathcal{H}\,,
\end{equation}
where $\mathrm{id}$ denote the identity operator on $\mathcal{E}$. We have the following uniform bound for the standard induced operator norm on $\mathcal{E}$:
\begin{equation}
\Vert R\Vert \leq \cos^{-1}(\varphi/2)\,.
\end{equation}
\end{lemma}
Our aim is to use this Lemma to bound each term in the tree expansion of $F$. We define:
\begin{equation}
\mathcal{B}_{\mathcal{T}_n}:=\prod_{\ell\in\mathcal{T}_n}\varpi_{I_{t(\ell)} I_{s(\ell)}} \prod_{v\in\mathcal{T}_n} \mathcal{V}_{I_{v,1},\cdots,I_{v,c(v)}}(\chi_p)\,.
\end{equation}
Let us consider a tree $\mathcal{T}$ and an edge $\ell\in\mathcal{T}$ with boundary $\partial \ell =\{v,w\}$. Let us denote as $V_I^{(v)}$ and $\bar{V}_I^{(w)}$ the two connected components connected together with the bridge $\ell$, so that the amplitude $\mathcal{A}_{\mathcal{T}}$ reads:
\begin{equation}
\mathcal{B}_{\mathcal{T}}=\sum_{I,J} V_I^{(v)} \mathcal{W}_{IJ}\bar{V}_J^{(w)}= \sum_L \mathcal{O}(V)_L^{(v)} (\lambda_L)^2\mathcal{O}(\bar{V})_L^{(w)}\,, \label{amplitudeutile}
\end{equation}
where $\mathcal{O}(V)_L^{(v)}:= \sum_I V_I^{(v)}O_{IL}$. Denoting as $\lambda_0^2\leq \sum_I \mathcal{W}_{II}$ the highest eigenvalue $\lambda_L$, we get:
\begin{equation}
\mathcal{B}_{\mathcal{T}}\leq (\lambda_0)^2 \sum_L \mathcal{O}(V)_L^{(v)} \mathcal{O}(\bar{V})_L^{(w)} = (\lambda_0)^2\sum_L V_L^{(v)} \bar{V}_L^{(w)} \,.
\end{equation}
Finally, using the standard Cauchy-Schwarz inequality:
\begin{align}
\nonumber \left\vert \sum_L V_L^{(v)} \bar{V}_L^{(w)} \right\vert \leq {\left(\sum_I \vert V_I^{(v)}\vert \right)\left(\sum_J \vert V_J^{(w)}\vert\right)}
\end{align}
recursively for all edges of the tree, we get the first bound:
\begin{equation}
\mathcal{B}_{\mathcal{T}_n}\leq(\lambda_0)^{2(n-1)} \prod_{v\in\mathcal{T}_n} \frac{1}{2}\Tr \vert R^{c(v)}(\chi_p) \vert\,.\label{equationvertex}
\end{equation}
From Lemma \ref{lemmauseful2}, we have $\Tr \vert R^{c(v)}(\chi_p) \vert \leq \left\vert\cos \frac{\varphi}{2} \right\vert^{-c(v)} \Tr\,\textbf{I}$, where $\textbf{I}$ is the $N\times N$ identity matrix. We thus get the second bound:
\begin{equation}
\mathcal{B}_{\mathcal{T}_n}\leq\frac{(\lambda_0)^{2(n-1)} }{2^n}\left\vert\frac{1}{\cos^2\frac{\varphi}{2} }\right\vert^{n-1} \left(\Tr\, \textbf{I}\right)^n =\frac{(\lambda_0)^{2(n-1)} }{2^n} \left\vert\frac{1}{\cos^2\frac{\varphi}{2} }\right\vert^{n-1}\, N^n
\end{equation}
where we used the topological relation $\sum_v c(v)=2(n-1)$. From Cayley formula, we easily deduce that the number of labelled trees with $n$ vertices for a fixed configuration for coordination numbers $\{c(v)\}$ is bounded by $\Omega(n,\{c(v)\})\leq\frac{n!}{\prod_v (c(v)-1)!}$, and the remaining sum over coordination numbers is trivially bounded with the area of the $(n-1)$-dimensional sphere of radius $\sqrt{2(n-1)}$, leading to:
\begin{equation}
\frac{1}{n!}\vert F_n\vert \leq \sqrt{\frac{2}{\pi}}\frac{24e^2 \vert\cos(\varphi/2)\vert^2}{u\,\lambda_0^2}\, \left\vert\frac{N\,\lambda_0^2}{\cos^2\frac{\varphi}{2} }\frac{u}{6}\sqrt{\frac{\pi}{e}}\right\vert^{n}\,,
\end{equation}
where we used of the Stirling formula. Finally, for $F_0$, a simple integration by parts leads to:
\begin{equation}
\begin{aligned}
&\left\vert\int \droit\mu_{\varpi\otimes x}(\chi^{(I)}) \Tr\ln\left(1-\sqrt{\frac{{-u}}{3} } \chi\right)\right\vert \leq (\lambda_0)^2 \left\vert\int \droit\mu_{\varpi\otimes x}(\chi^{(I)})\int_0^1 \droit t \, \frac{\frac{\rho}{3}}{(1-\sqrt{\frac{{-u}}{3} } \chi t)^2}\right\vert\,,
\end{aligned}
\end{equation}
which is trivially bounded. Thus, we explicitly showed the existence of a finite analyticity domain, the interior of the cardioid $\rho \leq \frac{6}{N\, \lambda_0^2}\sqrt{\frac{e}{\pi}} \cos^2 \left(\frac{\varphi}{2}\right)$.

\begin{remark}
\label{rk:0d-boson-K}
\emph{For the derivation of the bound, we assumed the i.i.d Gaussian measure $K_{ij}=\delta_{ij}$. This condition can be easily relaxed. First of all, Lemma \ref{lemmauseful2} still holds and the resolvent $R(\chi)$ remains bounded as $\cos^{-1}(\varphi/2)$. Each effective vertex involves one $K^{-1}$ insertion per derivative with respect to the intermediate field, replacing $\Tr \textbf{I}\to \Tr K^{-c(v)}$.}
\end{remark}

Now, we move on to the relativistic model in $d=1$. Without lost of generality, we consider a complex model $\phi_i(x)\in \mathbb{C}$ at finite temperature $\beta$, such that the field can be expanded in Fourier modes as $\phi_i=\frac{1}{\sqrt{\beta}}\sum_{n\in\mathbb{Z}} \varphi_i(n) \, \exp \left(i\frac{2\pi n}{\beta} t\right)$. The propagator of the theory is then $C(\omega,\omega^\prime)=C(\omega)\delta(\omega-\omega^\prime)$, with:
\begin{equation}
C(\omega)=\frac{1}{\omega^2+m^2}\,.
\end{equation}
As in the previous case, we introduce $\Psi_L:=\sum_{I=ij}O_{IL}\phi_i\bar{\phi}_j$, and break the quartic interaction $\int dt \vert \Psi_L\vert^2$ introducing a time-dependent intermediate field $\sigma^{(L)}$ with normalized Gaussian measure $\droit\nu(\sigma)$:
\begin{equation}
Z(u)=\int \droit\mu_C(\phi,\bar{\phi})\droit\nu(\sigma) \prod_{L}e^{\sqrt{\frac{{-u}}{2} }\lambda_L \int \droit t \Psi_L\sigma^{(L)}}\,.
\end{equation}
Integrating over the complex field $\Phi$, we get an effective matrix-field theory:
\begin{equation}
Z(u)=\int \droit\nu(\sigma) \, e^{- {\textbf{Tr}}\ln\mathcal{K}[\sigma]}\,,\label{equationnewZ}
\end{equation}
where the big trace $\textbf {Tr}$ sums over internal indices and frequencies and $\mathcal{K}$ is the operator with entries ($\mathcal{W}^{\frac{1}{2}}_{ijkl}:= O_{ij,mn} \lambda_{mn}O_{mn,kl}^T$):
\begin{equation}
\mathcal{K}_{ij}(\omega,\omega^\prime):=\left(\delta_{ij}\delta_{\omega\omega^{\prime}}-\sqrt{\frac{-u}{2\beta}} \frac{1}{\omega^2+m^2} \sum_{kl} \mathcal{W}^{\frac{1}{2}}_{ijkl} \sigma_{kl}(\omega-\omega^\prime)\right)\,.
\end{equation}
The only changes come from the dependence on $\omega$ of the intermediate fields, as well as the additional bound coming from $\omega$-integrations. Introducing the fields $\chi(\omega)$ like in \eqref{R}, the new resolvent is:
\begin{equation}
R[\chi](\omega,\omega^\prime):=\mathcal{K}^{-1}[\chi](\omega,\omega^\prime)\,.\label{resolvent2}
\end{equation}
Note that the operator $C\Sigma$, with $\Sigma$ having entries $\Sigma_{ij,\omega\omega^\prime}:= \sum_{kl}\mathcal{W}^{\frac{1}{2}}_{ijkl}\sigma_{kl}(\omega-\omega^\prime)$, is not Hermitian. However, this operator appearing in the trace, we can replace it by $C^{1/2}\Sigma C^{1/2}$, which is obviously Hermitian. The resolvent is thus bounded by $\cos^{-1}(\varphi/2)$ thanks to the Lemma \ref{lemmauseful2}. Defining $F_n=\frac{(-1)^n}{2^{n-1}}\sum_{\mathcal{T}_n} \mathcal{A}_{\mathcal{T}_n}$, where $\mathcal{A}_{\mathcal{T}_n}$ denotes the amplitude of the tree $\mathcal{T}_n$ with $n$ vertices, we have the following Lemma:
\begin{lemma}
The amplitude $\mathcal{A}_{\mathcal{T}_n}$ for the amplitude associated to the tree $\mathcal{T}_n$ with $n$ vertices admits the following bound:
\begin{equation}
\vert \mathcal{A}_{\mathcal{T}_n}\vert \leq \left\vert\frac{u \,\lambda_0^2}{4\beta} \right\vert^{n-1} \frac{(\Tr\textbf{I})^n }{\cos^{2n-2}(\varphi/2)} \,\left(\frac{\beta}{2m }\right)^{2n-2}\,\coth^{2n-2}\left(\frac{\beta m}{2}\right)\,\sum_{\omega} 1\,.
\end{equation}
\end{lemma}
\textit{Proof.} This statement can be easily proved recursively in the number of vertex. Let us consider the case $n=2$. The amplitude reads explicitly:
\begin{equation}
\mathcal{A}_{\mathcal{T}_2}=\frac{1}{2}\sum_\omega \left(\int_0^1 \droit x\right)\int \droit\nu_{X}[\chi] \,\frac{\partial}{\partial \chi^{(I)}_{1}(\omega)} \mathcal{W}_{IJ} \frac{\partial}{\partial\chi^{(J)}_{2}(-\omega)} \prod_{p=1}^2\mathcal{V}(\chi_p)\Big\vert_{\chi^{(L)}_p=0}\,,
\end{equation}
where the Gaussian measure is for the propagator $X_{ij}^{IJ}(\omega,\omega^\prime)=\mathcal{W}_{IJ} x_{ij} \delta_{\omega,-\omega^\prime}$. Computing the derivatives with respect to $\chi^{(J)}_{2}$ and $\chi^{(I)}_{1}$, we get, for each of them:
\begin{align}
\nonumber \frac{\partial}{\partial\chi^{(I)}_{1}(\omega)} \mathcal{V}(\chi_1)&=-\sum_{\omega^{\prime} \omega^{\prime\prime}}\sqrt{\frac{-u}{2\beta}}\, \frac{1}{\omega^{\prime\prime\,2}+m^2} \, \delta(\omega-\omega^{\prime\prime}+\omega^\prime)\mathcal{V}_I[\chi_1](\omega^\prime,\omega^{\prime\prime})\,.
\end{align}
Formally, the last term has the structure of a trace over $\omega$, $-\sqrt{\frac{-u}{2\beta}} \Tr B\mathcal{V}_{I}[\chi_1]$, involving the strictly positive matrix
\begin{equation}
B_{\omega^\prime\omega^{\prime\prime}}(\omega):=\frac{1}{\omega^{\prime\prime\,2}+m^2} \, \delta(\omega-\omega^{\prime\prime}+\omega^\prime).
\end{equation}
For $A$ bounded and $B$ strictly positive $\vert \Tr(AB)\vert \leq \Vert A \Vert \Tr B$. However, $B$ is not diagonalizable. In fact, only the $k$th diagonal of the matrix is non-vanishing (counting from the principal diagonal): $B_{ij}=b_i \delta_{i,j-k}$ with $b_i>0$ and $k \neq 0$. For our purpose, $A\equiv R$, and the resolvent is unitary diagonalizable as a function of the Hermitian operator $H$. Denoting as ${U}$ the unitary operator diagonalizing $A$ and as $a_i$ its eigenvalues, we get:
\begin{equation}
\vert \Tr(AB)\vert \leq \vert \sum_{l,i} a_l \, {U}_{i-k,l} {U}_{l,i}^\dagger b_i \vert \leq \Vert A\Vert \sum_{l,i} \vert {U}_{i-k,l} {U}_{l,i}^\dagger b_i \vert\,.
\end{equation}
Using Cauchy-Schwarz inequality, the last product can be bounded as:
\begin{equation}
\sum_{i,l}\vert {U}_{i-k,l} {U}_{l,i}^\dagger b_i \vert = \sum_i b_i \sum_l \vert {U}_{i-k,l} \vert \vert {U}_{i,l}\vert \leq \sum_i b_i \sqrt{\sum_{l}\vert {U}_{i-k,l}\vert^2\sum_{l}\vert{U}_{i,l}\vert^2}\,,
\end{equation}
that is to say, because ${U} {U}^\dagger = \mathrm{id}$: $\vert \Tr(AB)\vert \leq \Vert A \Vert \,\left(\sum_i b_i\right)$, and we get:\footnote{We used the well known sum $\sum_{n\in\mathbb{Z}}\, \frac{1}{n^2+y^2}=\frac{\pi}{y}\coth(\pi y)$.}
\begin{equation}
\sum_I\left\vert\frac{\partial}{\partial\chi^{(I)}_{1}(\omega)} \mathcal{V}(\chi_1)\right\vert \leq \frac{1}{\cos(\varphi/2)} \left\vert \frac{u}{2\beta} \right\vert^{1/2}(\Tr\textbf{I})\,\frac{\beta}{2m}\coth\left(\frac{\beta m}{2}\right)\,,
\end{equation}
where we used Lemma \ref{lemmauseful2}. Similarly to the $d=0$ case, from the Cauchy-Schwarz inequality and taking into account the normalization of the integrals, we get:
\begin{equation}
\vert \mathcal{A}_{\mathcal{T}_2} \vert \leq \,\left\vert\frac{u \,\lambda_0^2}{16} \right\vert \frac{(\Tr\,\textbf{I})^2}{\cos^2(\varphi/2)}\,\frac{\beta}{m^2} \coth^2\left(\frac{\beta m}{2}\right)\, \sum_\omega 1 \,.\label{stepA1}
\end{equation}
Now, we assume that the statement holds for trees having $n-1$ vertices. Any tree with $n$ vertices can be obtained from a tree with $n-1$ vertices by adding a leaf. Up to Gaussian and $x$ integrations, we get the explicit expression for the amplitude $\mathcal{A}_{\mathcal{T}_n}$:
\begin{equation}
\mathcal{A}_{\mathcal{T}_n} \propto \frac{1}{2^{n}}\left(\frac{u}{2\beta}\right)^n \sum_{\omega }\, \mathcal{R}_{K,\omega}\mathcal{W}_{KL} \mathcal{U}_{L;\,\omega }\,,
\end{equation}
$\mathcal{R}_{K,\omega}$ denoting the rest of the diagram, up to the added leaf, the Cauchy-Schwartz inequality leads to the bound:
\begin{equation}
\vert \sum_{\omega }\, \mathcal{R}_{K,\omega}\mathcal{W}_{KL} \mathcal{U}_{L;\,\omega } \vert \leq \lambda_0^2 \Vert \mathcal{R}_{K,\omega} \Vert \Vert \mathcal{U}_{L;\,\omega } \Vert (\Tr\textbf{I}) \sum_{\omega} \,1\,. \label{ineq1}
\end{equation}
We introduced the notation:
\begin{equation}
\mathcal{U}_{I_1,\cdot,I_n;\,\omega_1,\cdots,\,\omega_n}:=\left(\sqrt{\frac{3\beta}{-u}}\right)^n\left[\prod_{i=1}^n\frac{\partial}{\partial\chi_{I_i}(\epsilon_i\omega_i)}\right] \mathcal{V}(\chi_p)\,,
\end{equation}
where $\epsilon_i=\pm$. Explicitly, the structure of this vertex is the following:
\begin{equation}
\mathcal{U}_{I_1,\cdot,I_n;\,\omega_1,\cdots,\,\omega_n}= R_{j_1i_2;\,\bar{\omega}_1\bar{\omega}^\prime_1}(C\chi)_{\bar{\omega}_1^{\prime}\bar{\omega}_2;I_1\omega_1}R_{j_2i_3;\,\bar{\omega}_2\bar{\omega}^\prime_2}\cdots R_{j_{n}i_1;\,\bar{\omega}_n\bar{\omega}_{n}^{\prime}}(C\chi)_{\bar{\omega}_{n}^{\prime}\bar{\omega}_1;I_n\omega_n}
\end{equation}
where we integrated over repeated $\omega$-indices, and explicitly:
\begin{equation}
(C\chi)_{\bar{\omega}_i^{\prime}\bar{\omega}_{i+1};I_i\omega_i} =\frac{1}{\bar{\omega}_i^{\prime\,2}+m^2}\delta(\epsilon_i\omega_i-\bar{\omega}_i^{\prime}+\bar{\omega}_{i+1})\,.
\end{equation}
The effective vertex can be easily bounded using successive Cauchy-Schwartz inequalities as:
\begin{equation}
\sum_{I_1,\cdots,I_k} \vert \mathcal{U}_{I_1,\cdots,I_k;\,\omega_1,\cdots,\omega_k} \vert \leq \frac{\Tr\textbf{I}}{\cos^k(\varphi/2)} \delta\left(\sum_{i=1}^k \epsilon_i\omega_i\right)\,\left(\frac{\beta}{2m }\right)^{k}\,\coth^{k}\left(\frac{\beta m}{2}\right)\,.
\end{equation}
Therefore, from the recursion hypothesis, the positivity of the propagator and the momentum conservation along the loops ensured by the Dirac delta, we must have for the one-point amplitude $ \mathcal{R}_{K,\omega}$:
\begin{equation}
\vert \mathcal{R}_{K,\omega} \vert \leq \left\vert\frac{u \,\lambda_0^2}{4\beta} \right\vert^{n-2} \frac{(\Tr\textbf{I})^{n-1} }{\cos^{2n-4}(\varphi/2)} \,\left(\frac{\beta}{2m }\right)^{2n-4}\,\coth^{2n-4}\left(\frac{\beta m}{2}\right)\,.
\end{equation}
The statement following from inequality \eqref{ineq1}.
\begin{flushright}
$\square$
\end{flushright}
Thus, the series is absolutely convergent in the interior of the cardioid domain:
\begin{equation}
\label{eq:domain-1d-r-scalar-finite-T}
\rho \leq \frac{16\sqrt{e/\pi}\,m^2}{N\lambda_0^2\beta\coth^2\left(\frac{\beta m}{2}\right)} \cos^2 \left(\frac{\varphi}{2}\right)\,.
\end{equation}
Note that the massless limit of \eqref{eq:domain-1d-r-scalar-finite-T} is ill-defined because the RHS goes to zero. Moreover, in the continuum limit $\beta\to\infty$ (zero-temperature limit), the bound still exists. Indeed, in the low temperature limit, $t \in \mathbb R$, and the sums over $\omega$ become integrations:
\begin{equation}
\sum_\omega f(\omega) \to \frac{\beta}{2\pi} \int \droit \omega \, f(\omega)\,.
\end{equation}
In that limit, $\beta$ completely disappears, and the Fourier series become continuous Fourier transformations:
\begin{equation}
\phi_i=\frac{1}{\sqrt{2\pi}}\int_{-\infty}^{+\infty} \droit\omega \, \varphi_i(\omega)e^{i\omega t}\,,
\end{equation}
and from the integral $\int_{-\infty}^{+\infty} \frac{\droit \omega}{\omega^2+m^2}=\frac{\pi}{m}$, the rule to pass from finite to infinite $\beta$ is to replace $\beta$ by $2\pi$ everywhere, and set $\coth(\beta m/2)=1$. Thus the cardioid domain becomes:
\begin{equation}
\rho \leq \frac{8 \sqrt{e/\pi^3}\, m^2}{N\lambda_0^2} \cos^2 \left(\frac{\varphi}{2}\right) \,.
\end{equation}

\subsubsection{Borel summability} \label{secBorel}

The last piece to prove Borel summability is to check the second requirement of Nevanlinna's theorem. The proof is classic and follows immediately from the previous results, and we provide only the main steps in $d=0$, the extension to $d=1$ being straightforward.

The remainder of the expansion of the free energy reads ($g=u/4!$):
\begin{equation}\label{rest2}
R_rF'(g):=g^{r+1}\int_{0}^1\frac{(1-t)^r}{r!}(F')^{(r+1)}(tg)\droit t\,,
\end{equation}
with $F^\prime:= \sum_{n=1}^\infty F_n=F-F_0$. The remainder can be expanded as a sum over labelled trees thanks to the BKAR formula, and we get:
\begin{equation}\label{restF}
R_rF'(g):=-\sum_{n=1}^{\infty}\frac{1}{n!}\sum_{\mathcal{T}_n}\prod_{\ell\in\mathcal{T}_n}\left(\int_0^1 \droit x_{\ell}\right) \int \droit\mu_{\varpi\otimes x}(\chi^{(I)}_p)R_r[\mathcal{Y}_{\mathcal{T}_n}]\,,
\end{equation}
where
\begin{equation}\label{Y}
\mathcal{Y}_{\mathcal{T}_n}:=g^{n-1}\prod_{v\in\mathcal{T}_n} (c(v)-1)!
\prod_{\ell\in\mathcal{T}_n}\varpi_{I_{t(\ell)} I_{s(\ell)}} \prod_{v\in\mathcal{T}_n} \mathcal{V}_{I_{v,1},\cdots,I_{v,c(v)}}(\chi_p)\Big\vert_{\chi^{(L)}_p=0, x_{pq}=1}\,.
\end{equation}
First, note that when $n-2\geq r$, $R_r[\mathcal{Y}_{\mathcal{T}_n}]=\mathcal{Y}_{\mathcal{T}_n}$. Each of these terms admits a bound of the form $K^{n}|\lambda|^n$, and the sum converges at least in the interior of the cardioid. When $n-2<r$, however, the remainder has to be computed following a Taylor expansion of the resolvents in $\mathcal{Y}_{\mathcal{T}}$. This expansion dresses the trees with loops. Concretely, we extract the global factor $g^{n-1}$ in front of $\mathcal{Y}_{\mathcal{T}}$, and we write:
\begin{equation}
\mathcal{Y}_{\mathcal{T}_n}=g^{n-1}\bar{\mathcal{Y}}_{\mathcal{T}_n}\,,
\end{equation}
such that:
\begin{equation}
R_r[\mathcal{Y}_{\mathcal{T}_n}]=g^{n-1}R_{r-n+1}[\bar{\mathcal{Y}}_{\mathcal{T}_n}]\,.
\end{equation}
Defining $z:= \sqrt{-u/3}$, we have:
\begin{equation}
\frac{\droit}{\droit z}\Tr R(\chi)=n! \Tr [\chi^n R^{n+1}(\chi)]\,.
\end{equation}
As announced, the tree $\mathcal{T}_n$ becomes a spanning tree for a graph with loops, generated from the new Gaussian integrations of the fields $\chi_v$ coming from the Taylor expansion of the resolvent. Each power of the resolvent can be bounded using Lemma \ref{lemmauseful2}, and we deduce:
\begin{align}\label{sumtree2}
\nonumber R_{r-n+1}[\bar{\mathcal{Y}}_{\mathcal{T}_n}]\leq \frac{ z^{2(r-n)}}{2^n}&\int_{0}^1\droit t \, \frac{(1-t)^{2r-2n+2}}{(2r-2n+2)!} \sum_{\{k_l\}|\sum_{k_l}=2r-2n+3}\frac{(2r-2n+3)!}{\prod_{l=1}^n k_l!}\\
&\times\prod_{v=1}^n(c(v)+k_v-1)! \times \Tr\left[ \textbf{I} \prod_{v\in\mathcal{T}_n} \chi_v^{k_v}\right] \Big\vert_{\chi^{(L)}_p=0} \,.
\end{align}
We can now plug this expression in equation \eqref{restF}. First, note that because $C_n^p\leq 2^n$, the factor $c(m)(c(m)+k_m)!/[k_m!c(m)!]$ is bounded by $e^{\ln(c(m))}2^{c(m)+k_m}\leq (2e)^{c(m)}2^{k_m},$ and the product over $m$ gives a factor $(2e)^{2n-2}2^{2r-2n+3}$. Secondly, we can perform the Gaussian integration. Because there are $2r-2n$ fields, the number of Wick contractions is $(2(r-n)!! = 2^{r-n}(r-n)!\leq 2^rr!$. Finally, the remaining integration over $t$ gives $(2r-2n+3)^{-1}$, which together with the denominator $(2r-2n+2)!$ exactly compensates the combinatorial factor $(2r-2n+3)!$. As in the previous section, using Cayley's theorem for the number of trees with $n$ vertices and Stirling's formula, we find a bound of the form: $AB_1^nB_2^rr!$ for some constants $A$, $B_1$ and $B_2$. Because $n-2<r$, summing over $n$, we find the final bound: $A'|g|^rB^rr!$ for the contributions in \eqref{restF} for which $n-2<r$. As explained before, the contributions for $n-2\geq r$ are all bounded by bounds of the form: $|g|^{n}K^{n}$, and the sum behaves as: $A''|g|^rK^r$. Ultimately, because, for positive constants $k_1$ and $k_2$, $k_1r!+k_2\leq (k_1+k_2)r!$, we find that $|R_rF(g)|\leq A''(B')^r|g|^rr!$.

\subsection{Bosons in dimension \texorpdfstring{$d = 2$}{d = 2}}

In this section, we prove the existence of a finite analytic domain with a cardioid shape. It requires improving the standard LVE to deal with UV divergences, which is achieved by the so-called MLVE~\cite{Gurau:2015:MultiscaleLoopVertex, Gurau:2014:RenormalizationAdvancedOverview}.
It allows us to subtract scale by scale the divergent subgraphs, from an expansion requiring “forests into forests”, and known as \emph{jungle expansion}.

\subsubsection{Renormalized model and jungle expansion}

From the power-counting theorem  \eqref{crit2}, the degree of divergence for two-dimensional relativistic bosons becomes independent of the number of external edges $N$, and show that divergences arises only from diagrams with a single vertex, that is for $(V,N)=(1,2)$ and $(V,N)=(1,0)$, and are both logarithmic. We only consider the case of complex bosons. From an elementary perturbative calculation using Feynman rules for the interaction:
\begin{equation}
\mathcal{S}_{\text{int}}:=\frac{u}{4!} \sum_{ijkl}\int_{-\beta/2}^{+\beta/2} \droit t\int_{-\infty}^{+\infty} \droit x \, \mathcal{W}_{ijkl} \phi_i(x,t)\bar{\phi}_j(x,t)\phi_k(x,t)\bar{\phi}_l(x,t)\,,
\end{equation}
we get the explicit expressions of the two divergent diagrams:
\begin{enumerate}
\item The vacuum diagram $(V,N)=(1,0)$:
\begin{equation}
\vcenter{\hbox{\includegraphics[scale=1]{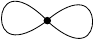} }}=-\frac{u}{2\pi\beta}\left( \sum_\omega\int_{-\infty}^{+\infty} \frac{\droit p}{\omega^2+p^2+m^2}\right)^2 \sum_{ij} \mathcal{W}_{iijj} \,,\label{div1}
\end{equation}

\item The $2$-point diagram $(V,N)=(1,2)$:
\begin{equation}
\vcenter{\hbox{\includegraphics[scale=1]{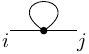} }}=-\frac{u}{\pi\beta}\left( \sum_\omega\int_{-\infty}^{+\infty} \frac{\droit p}{\omega^2+p^2+m^2}\right) \sum_{k} \mathcal{W}_{ijkk}\,.\label{div2}
\end{equation}
\end{enumerate}
Both of these diagrams scale as $\ln(\Lambda)$, with $\Lambda$ designating some arbitrary UV cut-off in the sums and integrals. A renormalized (i.e. UV finite) action can be obtained including two counter-terms in order to systematically subtract the divergent diagrams. The two counter-terms corresponding respectively to $2$-point and vacuum divergences are denoted by $\delta V_1$ and $\delta V_2$ and are respectively given by:
\begin{equation}
\delta V_1(\bar{\Phi},\Phi):= \sum_\omega\int_{-\infty}^{+\infty} \droit p\, \bar{\phi}_i(\omega,p)\delta m^2_{ij}\phi_j(\omega,p)\,,
\end{equation}
with $\delta m^2_{ij}=:\Delta^2\sum_{k} \mathcal{W}_{jikk}$, and:
\begin{equation}
\Delta^2:=-\frac{u}{\pi\beta}\, \sum_\omega\int_{-\infty}^{+\infty} \frac{\droit p}{\omega^2+p^2+m^2}\,,
\end{equation}
and
\begin{equation}
\delta V_2:=-\frac{u}{2\pi\beta}\left( \sum_\omega\int_{-\infty}^{+\infty} \frac{\droit p}{\omega^2+p^2+m^2}\right)^2 \sum_{ij} \mathcal{W}_{iijj}\,.
\end{equation}
The mass divergence arising from the tadpole can be subtracted by adding $\delta V_1$ to the original action. However, we have to be careful, because this counter-term generates a new vacuum divergence as well. Nonetheless, it is easy to check that this new divergence can be subtracted by a counter-term equals to $-2\delta V_2$. Therefore, the cancellation of all vacuum divergences requires the counter-term $\delta Vac= \delta V_2-2\delta V_2= - \delta V_2$. Thus, the completely renormalized model reads:
\begin{equation}
Z_{\text{R}}(J,\bar{J})=e^{\delta V_2}\int \droit\Phi \droit\bar{\Phi} e^{-\mathcal{S}_{\text{PR}}(\bar{\Phi},\Phi)+\sum_i \bar{j}_i\phi_i+\sum_i\bar{\phi}_i j_i}\,,\label{fullrengenerating}
\end{equation}
with:
\begin{equation*}
S_{\text{int,PR}}(\bar{\Phi},\Phi)=\frac{1}{2}\frac{u}{2\pi \beta} \sum \int \mathcal{W}_{ijkl}\phi_i\bar{\phi}_j\phi_k\bar{\phi}_l +\Delta^2\sum\int \left(\sum_k\mathcal{W}_{ijkk}\right) \phi_i \bar{\phi}_j\,.
\end{equation*}
Completing the square, we get:
\begin{equation}
S_{\text{int}}(\bar{\Phi},\Phi)=\frac{1}{2}\frac{u}{2\pi \beta} \sum_I \left(\mathcal{M}_I+\frac{2\pi \beta \Delta^2 }{u} \mathcal{X}_I \right)^2-\frac{\pi\beta}{2u}(\Delta^2)^2\sum_I \mathcal{X}_I^2\,,
\end{equation}
where we defined $\mathcal{M}_{kl}:=\sum_{i,j}\mathcal{W}^{\frac{1}{2}}_{ijkl}\phi_i \bar{\phi}_j$ and $\mathcal{X}_{ij}:= \sum_k\mathcal{W}^{\frac{1}{2}}_{ijkk} $, and the last term is explicitly:
\begin{equation}
\frac{\pi\beta}{2u}(\Delta^2)^2\sum_I \mathcal{X}_I^2=\frac{u}{2\pi\beta} \left( \sum_\omega\int_{-\infty}^{+\infty} \frac{\droit p}{\omega^2+p^2+m^2}\right)^2 \sum_{ij} \mathcal{W}_{iijj}=-\delta V_2\,.
\end{equation}
Next, introducing intermediate fields to break the square, integrating over the fields $\Phi$ and $\bar{\Phi}$, and rewriting the determinant as a $\Tr$-$\log$ effective interaction, we finally get the effective model for the intermediate fields $\sigma=\{\sigma_I\}$:
\begin{equation}
Z(u)=\int \droit \nu(\sigma) e^{-\textbf{Tr}\ln\mathcal{K}[\sigma]-\sqrt{\frac{2\pi\beta}{-u}} \Delta^2 \int \sum_I \mathcal{X}_I \sigma_I }\,, \label{intermediatepart2}
\end{equation}
where $\mathcal{K}$ is the matrix with elements:
\begin{equation}
\mathcal{K}_{ij}(\textbf{p},\textbf{p}^\prime):=\delta_{ij}\delta(\textbf{p},\textbf{p}^\prime)-\sqrt{\frac{-u}{2\pi\beta}} \frac{1}{\textbf{p}^2+m^2}\mathcal{W}^{\frac{1}{2}}_{ijkl}\sigma_{kl}(\textbf{p}-\textbf{p}^\prime)\,, \label{newK}
\end{equation}
with $\delta(\textbf{p},\textbf{p}^\prime):=\delta_{\omega\omega^\prime}\delta(p-p^\prime)$ and $\textbf{Tr}$ in formula \eqref{intermediatepart2} means
\begin{equation}
\textbf{Tr} \, \mathcal{K}:=\sum_{i}\int \droit \textbf{p} \,\mathcal{K}_{ii}(\textbf{p},\textbf{p})\,,
\end{equation}
with the shorthand notation $\int \droit \textbf{p}:=\sum_\omega\int \droit p$. Note that the linear terms in $\sigma$ in the equation \eqref{intermediatepart2} exactly compensate the first term in the Taylor expansion of the logarithm, in such a way that the effective partition function may be rewritten as:
\begin{equation}
Z(u)=\int \droit \nu (\sigma) \, e^{-\textbf{Tr}\ln_2\mathcal{K}[\sigma] }\,. \label{intermediatepart3}
\end{equation}
We recall the standard definition $\ln_2(1-x)=\ln(1-x)+x$. The notation $\ln_2$ in this expression makes sense because $\mathcal{K}$ has the form $\mathcal{K}=:\mathrm{Id}/\mathcal{H}$, the explicit expression for the operator $\mathcal{H}$ being given on equation \eqref{newK}. Defining the resolvent $R=\mathcal{K}^{-1}$, we have once again the resolvent bound:
\begin{equation}
\Vert R \Vert \leq \cos^{-1}\left(\frac{\varphi}{2}\right)\label{boundresolventmlve}
\end{equation}
where $\varphi=\arg(u)$, $\vert u \vert \leq \pi$. This bound follows from the Lemma \ref{lemmauseful2} and from the same arguments used in the case $d=1$ to transform $C\Sigma$ as $C^{1/2} \Sigma C^{1/2}$.

\medskip

Now, we move on to the jungle expansion. First, we fix our different conventions and notations. A natural choice, taking into account the $\SO(2)$ invariance of the Laplacian is to choose a disk: $0\leq \textbf{p}^2\leq \Lambda^2$, with the shorthand notations $\textbf{p}:=(\omega,p)$ and $\textbf{p}^2:=\omega^2+p^2$. We refer to "momentum space" as the set of $\textbf{p}$ into the disk of radius $\Lambda$. In order to use the multi-scale expansion, we need to introduce a slicing into the momentum space. To this end, we consider a pair of real and integer numbers $(M,j_{\text{max}})$ and assume that $\Lambda=M^{j_{\text{max}}}$. Intermediate scales are then defined as $M^j$ for $j\leq j_{\text{max}}$, and we introduce the slice function $\chi_j(\textbf{p})$ such that:
\begin{equation}
\chi_j(\textbf{p}):=\chi_{\leq j}(\textbf{p})-\chi_{\leq j-1}(\textbf{p})\qquad j\geq 2\,,
\end{equation}
where $\chi_{\leq j}(\textbf{p})$ denotes the step function $\chi_{\leq j}(\textbf{p}):=\varphi(M^{2j}-\textbf{p}^2)$. We define
\begin{equation}
V_{\leq j}:=\textbf{Tr} \ln_2\left(\mathrm{id}+\chi_{\leq j}\mathcal{H}\chi_{\leq j}\right)\,,
\end{equation}
and
\begin{equation}
V_j:=V_{\leq j}-V_{\leq j-1}\,,
\end{equation}
such that $\sum_j V_j=V\equiv \textbf{Tr} \ln_2\left(\mathrm{id}+\mathcal{H}\right)$. In terms of this slice decomposition, the partition function \eqref{intermediatepart3} reads:
\begin{equation}
Z(u)=\int \droit \nu(\sigma)\prod_{j=0}^{j_{\text{max}}} \, e^{-V_j[\sigma] }\,. \label{intermediatepart4}
\end{equation}
The first MLVE trick is to view the product as a determinant, and to rewrite it as a Grassmann integral. More precisely, defining $W_j:=e^{V_j}-1$, we may rewrite \eqref{intermediatepart4} as:
\begin{equation}
Z(u)=\int \droit \nu(\sigma) \int \prod_{j=0}^{j_{\text{max}}} \droit \mu_{1}(\bar{\chi}_j,\chi_j) e^{-\sum_{j=0}^{j_{\text{max}}} \bar{\chi}_jW_j \chi_j}\,,
\end{equation}
with the Gaussian Grassmann measure with identity kernel:
\begin{equation}
\droit \mu_{1}(\bar{\chi}_j,\chi_j):=\droit\bar{\chi}_j \droit \chi_j \exp\left(-\bar{\chi}_j\chi_j\right).
\end{equation}
We are now in the position to perform the MLVE of the model. We only reproduce the main step of the general method, referring to the standard reference~\cite{Gurau:2015:MultiscaleLoopVertex} for technical details. Let us define $S=[0,j_{\text{max}}]$ to be the set of scales and $\mathbb{I}_S$ the $S\times S$ identity matrix, allowing to rewrite the partition function $Z(u)$ in the compact form
\begin{equation}
Z(u)=\int \droit\nu_S \, e^{-W} \,``= "\, \sum_{n=0}^\infty \frac{1}{n!}\int \droit\nu_s\, (-W)^n\,,
\end{equation}
where $\droit\nu_s:= \droit \nu(\sigma) \droit \mu_{\mathbb{I}_s}(\{\bar{\chi}_j,\chi_j\})$. The quotes around the equality symbol refer to the illegal permutation of sum and Gaussian integrations. In fact, the aim of our procedure is to give a sense for this equality. As for the $d=0$ and $d=1$ cases, we introduce replicas for the vertices in the set $V:=\{1,\cdots,n\}$:
\begin{equation}
Z(u)=\sum_{n=0}^\infty \frac{1}{n!}\int \droit\nu_{S,V}\prod_{a=1}^n(-W_a)\,,
\end{equation}
in such a way that the vertex $W_a$ express in terms of bosonic intermediate fields $\sigma^{a}$, and the Gaussian measure $ \droit\nu_{S,V}$ becomes $ \droit\nu_{S,V}=\droit \nu_{\textbf{1}_V}(\sigma) \droit \mu_{\mathbb{I}_s}(\{\bar{\chi}_j,\chi_j\})$, the covariance becoming the $V\times V$ matrix $\textbf{1}_V$ with all entries equals to $1$. The MLVE trick for factorizing the integral is to use forest formula two times. The first one, over the bosonic intermediate field, follows in the same way as for the previous cases. We introduce the covariance $x_{ab}=x_{ba}$, with $x_{aa}=1$ between the bosonic replicas, and use the forest formula \eqref{BKAR} to write, in the derivative representation (holding as well for Grassmann fields):
\begin{align}
\nonumber Z(u)=\sum_{n=0}^{\infty} &\frac{1}{n!}\sum_{\mathcal{F}} \int_{0}^1 \left(\prod_{\ell\in\mathcal{F}} \droit x_\ell \right)\left[e^{\frac{1}{2}\sum_{a,b}^n X_{ab} (x_\ell)\frac{\partial}{\partial \sigma^a} \frac{\partial}{\partial \sigma^b} +\sum_{j=0}^{j_{\text{max}}} \frac{\partial}{\partial \bar{\chi}_j}\frac{\partial}{\partial {\chi}_j}} \right]\\
&\times \prod_{\ell\in\mathcal{F}} \left(\frac{\partial}{\partial \sigma^{a(\ell)}}\frac{\partial}{\partial \sigma^{b(\ell)}}\right)\prod_{a=1}^n \left(-\sum_{j=0}^{j_{\text{max}}} \bar{\chi}_j W_j(\sigma^a) \chi_j \right)\bigg\vert_{\sigma=\chi=\bar{\chi}}=0\,.
\end{align}
where $a(\ell)$ and $b(\ell)$ denote the end points of the edge $\ell$. The forest $\mathcal{F}$ defines a natural partition of the set $V$ into blocks building of its connected spanning trees. We call such a block as $\mathcal{B}$, and denote by $V/\mathcal{F}$ the reduced set building of bosonic spanning trees $\mathcal{B}$. Next, we use the same replica trick by assigning the fermionic fields to these blocks, such that $\chi_j\to\chi_j^{\mathcal{B}}$ with covariance $y_{\mathcal{B}\mathcal{B}^\prime}=y_{\mathcal{B}^\prime\mathcal{B}}$ and $y_{\mathcal{B}\mathcal{B}} = 1$. Following standard notations, we call $L_F$ a generic fermionic edge between two blocks $\mathcal{B}$ and $\mathcal{B}^\prime$. Using the forest formula a second time, we obtain an expansion indexed by \emph{two-level jungles} $\mathcal{J}$ rather than forests~\cite{Abdesselam:1995:TreesForestsJungles}:
\begin{equation}
Z(u)=\sum_{n=0}^\infty \frac{1}{n!} \sum_{\mathcal{J}} \sum_{j_1}\cdots\sum_{j_n} \int_{0}^1 \droit z_{\mathcal{J}} \int \droit \nu_{\mathcal{J}} \,\partial_{\mathcal{J}} \left[\prod_\mathcal{B}\prod_{a\in\mathcal{B}} \left(\chi_{j_a}^{\mathcal{B}} W_{j_a}(\sigma^a)\chi^{\mathcal{B}}_{j_a}\right)\right]\,,
\end{equation}
where:
\begin{itemize}
\item The two-level jungles $\mathcal{J}=(\mathcal{F}_B,\mathcal{F}_F)$ are ordered pairs of bosonic and fermionic disjoint forests of the set $V$, denoted as $\mathcal{F}_B$ and $\mathcal{F}_F$. We denote as $\ell_B$ and $L_F$ the bosonic and fermionic edges of the two components of the jungle $\mathcal{J}$. The notation $\ell$ being kept to denote generic edges of $\mathcal{J}$.

\item $\droit z_\mathcal{J}$ means integration from $0$ to $1$ over parameters $z_\ell = x_\ell$ for $\ell \in \mathcal{F}_B$ and $z_\ell = y_\ell$ for $\ell \in \mathcal{F}_F$, for each $\ell\in\mathcal{J}$.

\item $\partial_{\mathcal{J}}$ is a compact notation for:
\begin{equation}
\begin{aligned}
\partial_{\mathcal{J}}:= &\ \prod_{\ell_B=(a,b)\in\mathcal{F}_B} \left(\frac{\partial}{\partial \sigma^{a}}\frac{\partial}{\partial \sigma^{b}}\right)
\\
&\times \prod_{L_F=(p,q)\in\mathcal{F}_F} \delta_{j_pj_q}\left(\frac{\partial}{\partial \bar{\chi}^{\mathcal{B}(a)}_{j_a}}\frac{\partial}{\partial {\chi}^{\mathcal{B}(b)}_{j_b}}+\frac{\partial}{\partial \bar{\chi}^{\mathcal{B}(b)}_{j_b}}\frac{\partial}{\partial {\chi}^{\mathcal{B}(a)}_{j_a}}\right)\,.
\end{aligned}
\end{equation}

\item The Gaussian measure $\droit \nu_{\mathcal{J}}$ has covariance $X(x_\ell)\otimes \textbf{1}_S$ for bosons and $Y(x_\ell)\otimes \mathbb{I}_S$ for fermions, that is to say, in derivative representation:
\begin{equation}
\droit \nu_{\mathcal{J}}\equiv\left[e^{\frac{1}{2}\sum_{a,b}^n X_{ab} (x_\ell)\frac{\partial}{\partial \sigma^a} \frac{\partial}{\partial \sigma^b} +\sum_{\mathcal{B}\mathcal{B}^\prime} Y_{\mathcal{B}\mathcal{B}^\prime}(x_\ell)\sum_{a\in\mathcal{B},b\in\mathcal{B}^\prime} \delta_{j_aj_b}\frac{\partial}{\partial \bar{\chi}_{j_a}^{\mathcal{B}}}\frac{\partial}{\partial {\chi}_{j_b}^{\mathcal{B}^{\prime}}}} \right]_{\sigma=\chi=\bar{\chi}=0}\,.
\end{equation}

\item $X_{ab}(x_\ell)$ is the infimum of the parameters $x_\ell$ for the bosonic component of the jungle, in the unique path from $a$ to $b$. It is set to be equal to zero if the path does not exist, and to $1$ if $a=b$.

\item $Y_{\mathcal{B}\mathcal{B}^\prime}(y_\ell)$ is the infimum of the fermionic parameters along the fermionic edges $L_F$, between the blocks $\mathcal{B}$ and $\mathcal{B}^\prime$.
\end{itemize}

As for the cases discussed in the previous section, our aim is now to bound the amplitudes indexed with jungles. More precisely, we will prove the existence of a finite analyticity domain for the free energy $F(u)=\ln Z(u)$, expanding in terms of \emph{connected jungles}:

\begin{equation}
F(u)=\sum_{n=0}^\infty \frac{1}{n!} \sum_{\mathcal{J} \text{ trees}} \sum_{j_1}\cdots\sum_{j_n} \int_{0}^1 \droit z_{\mathcal{J}} \int \droit \nu_{\mathcal{J}} \,\partial_{\mathcal{J}} \left[\prod_\mathcal{B}\prod_{a\in\mathcal{B}} \left(\chi_{j_a}^{\mathcal{B}} W_{j_a}(\sigma^a)\bar{\chi}^{\mathcal{B}}_{j_a}\right)\right]\,.\label{multiscaleF}
\end{equation}

\subsubsection{Bounds and convergence}

In this section, we will bound the bosonic and fermionic contributions of the decomposition \eqref{multiscaleF}. The technical parts of the bounds are essentially the same as for the general treatment given in~\cite{Gurau:2015:MultiscaleLoopVertex}, and we refer to this paper for the technical subtleties, only indicating the main steps of the proof and focusing on the specificity of the model.

\subsubsection{Fermionic integrals}

The fermionic part of the expansion \eqref{multiscaleF} is exactly the same as in~\cite{Gurau:2015:MultiscaleLoopVertex}. Due to the standard properties of Grassmann integration, the Gaussian integration over these variables can be written as:
\begin{align}\label{inttrans}
\int \prod_{\mathcal{B}}\prod_{a\in\mathcal{B}}\droit\bar{\chi}_{j_a}^{\mathcal{B}}&\droit\chi_{j_a}^{\mathcal{B}} e^{-\sum_{a,b}^n\bar{\chi}_{j_a}^{\mathcal{B}(a)}\textbf{Y}_{ab}\chi_{j_a}^{\mathcal{B}(b)}}\prod_{L_F\in\mathcal{F}_F}\delta_{j_{a(L_F)}j_{b(L_F )}}\\
\nonumber&\qquad\times\bigg(\bar{\chi}_{j_{a(L_F)}}^{\mathcal{B}(a(L_F))}{\chi}_{j_{b(L_F)}}^{\mathcal{B}(b(L_F))}+\bar{\chi}_{j_{b(L_F)}}^{\mathcal{B}(b(L_F))}{\chi}_{j_{a(L_F)}}^{\mathcal{B}(a(L_F))}\bigg),
\end{align}
with $\textbf{Y}_{ab}:=Y_{\mathcal{B}(a)\mathcal{B}(b)}\delta_{j_aj_{b}}$. We thus define:
\begin{equation}\label{minor}
\textbf{Y}_{m_1,\ldots,m_k}^{p_1,\ldots,p_k}:=\int \prod_{\mathcal{B}}\prod_{a\in\mathcal{B}}\droit \bar{\chi}_{j_a}^{\mathcal{B}}\droit \chi_{j_a}^{\mathcal{B}} e^{-\sum_{a,b}\bar{\chi}_{j_a}^{\mathcal{B}(a)}\textbf{Y}_{ab}\chi_{j_{b}}^{\mathcal{B}(b)}}\prod_{r=1}^k\bar{\chi}_{j_{r}}^{\mathcal{B}(r)}{\chi}_{j_{r}}^{\mathcal{B}(r)},
\end{equation}
as the minor of the matrix $\textbf{Y}$, having the lines $p_1\cdots p_k$ and the columns $m_1\cdots m_k$ deleted.
Taking into account the \emph{hard core constraint inside each block}, meaning that the integral \eqref{inttrans} vanishes as soon as two vertices belong to the same bosonic block $\mathcal{B}$ with the same scale attribution, one can rewrite the equation \eqref{inttrans} as:
\begin{equation}
\bigg(\prod_{\mathcal{B}}\prod\limits_{\substack{a,b\in\mathcal{B}\\a\neq b}}(1-\delta_{j_{a}j_{b}})\bigg)\bigg(\prod_{L_F\in\mathcal{F}_F}\delta_{j_{a(L_F)}j_{b(L_F)}}\bigg)\bigg(\textbf{Y}_{a_1,\ldots,a_k}^{p_1,\ldots,p_k}+\textbf{Y}_{p_1,\ldots,m_k}^{m_1,\ldots,p_k}+\cdots+\textbf{Y}_{p_1,\ldots,p_k}^{m_1,\ldots,m_k}\bigg)
\end{equation}
where the sum runs over the $2^k$ ways to exchange the upper and lower indices, and $k:=|\mathcal{F}_F|$ is the cardinal of the fermionic forest, and the first product over factors $(1-\delta_{j_{a}j_{b}})$ implements the hard core constraint. To bound the fermionic contribution, we have the important lemma:
\begin{lemma}
Due to the positivity of the covariance $\textbf{Y}$, for any $\{m_i\}$ and $\{p_i\}$ the minor $\textbf{Y}_{m_1,\ldots,m_k}^{p_1,\ldots,p_k}$ defined in \eqref{minor} satisfies:
\begin{equation}
|\textbf{Y}_{m_1,\ldots,m_k}^{p_1,\ldots,p_k}|\leq 1.
\end{equation}
\end{lemma}
This lemma can be easily checked, and the proof may be found in~\cite{Gurau:2015:MultiscaleLoopVertex}.

\subsubsection{Bosonic integrals}

We now move on to the bosonic integrals. From the two-level trees decomposition of the free energy, it is clear that bosonic integrations factorizes over each block $\mathcal{B}$. As a result, we can only consider a single block $\mathcal{B}$. It involves the Gaussian integration:
\begin{equation}\label{bosonicint}
\int \droit\nu_{\mathcal{B}} F_{\mathcal{B}}(\sigma)=e^{\frac{1}{2}\sum_{a,b\in\mathbb{B}}X_{ab}(\mathit{w}_l)\frac{\partial}{\partial\sigma_a}\frac{\partial}{\partial \sigma_b}}F_{\mathcal{B}}(\sigma)\big|_{\sigma=0}
\end{equation}
with $F_{\mathcal{B}}(\sigma)$ defined as:
\begin{equation}\label{defF}
F_{\mathcal{B}}(\sigma)=\int \prod_{\ell_B\in \mathcal{B}} \droit \textbf{p}_{\ell_B} \bigg(\frac{\partial^2}{\partial \sigma_{a(\ell_B)}(-\textbf{p}_{\ell_B})\partial \sigma_{b(\ell_B)}(\textbf{p}_{\ell_B})}\bigg)\prod_{a\in\mathcal{B}}W_{j_a}(\sigma_a)\,,
\end{equation}
where we introduced the explicit momentum dependence and the integrations over momenta along each bosonic edge. The derivatives $\partial/\partial \sigma$ can be evaluated using the famous \emph{Fa\`a di Bruno} formula, which extends the standard derivation rule for composed functions. Computing the first derivative of the potential $V_j$, we get, for $k>0$:
\begin{align}
\nonumber\prod_{l=1}^k\partial_{\sigma_{a(l)}}(-V_j)=&\sum_{\pi} \textbf{Tr} \bigg(\underbrace{(\partial_{\sigma_{a(\pi(1))}}\mathcal{H})\chi_{\leq j}R_{\leq j}\cdots (\partial_{\sigma_{a(\pi(k))}}\mathcal{H})\chi_{\leq j}R_{\leq j}}_\textrm{k times}\\
&-\underbrace{(\partial_{\sigma_{a(\pi(1))}}\mathcal{H})\chi_{\leq j-1}R_{\leq j-1}\cdots (\partial_{\sigma_{a(\pi(k))}}\mathcal{H})\chi_{\leq j-1}R_{\leq j-1}}_\textrm{k times}\bigg)\,,
\label{FdB}
\end{align}
where:
\begin{equation}
R_{\leq j}:=(1+\chi_{\leq j} \mathcal{H}\chi_{\leq j})^{-1} \,,
\end{equation}
and the sum over $\pi$ runs over permutation of $k$ elements, up to cyclic permutations. Then, the $k$-th derivative of $W_j$ can be deduced from the Fa\`a di Bruno formula \eqref{FdB}. For $k>0$ we get:
\begin{equation}
\prod_{l=1}^k\partial_{\sigma_{a(l)}}(-W_j)=e^{-V_j}\sum\limits_{\substack{\{m_l\}\\\sum_{l\geq1}lm_l=k}}\frac{k!}{\prod_{l\geq1}m_l!(l!)^{m_l}}\prod_{l\geq 1}[\partial_{\sigma}^l(-V_j)]^{m_l}\,,
\end{equation}
where we used the compact notation $\prod_{l\geq 1}[\partial_{\sigma}^l(-V_j)]^{m_l}$ to represent a block of derivative of size $l$. In \eqref{defF}, we can rewrite the product as a product over the \emph{arcs} of the vertices:
\begin{equation}\label{defF2}
F_{\mathcal{B}}(\vec{\sigma})=\prod_{v\in\mathcal{B}}\left[\prod_{k=1}^{c(v)}\dfrac{\partial}{\partial \sigma_{a(v)}(\epsilon_k \textbf{p}_k)}\right]W_{j_v}(\sigma_{a(v)}),
\end{equation}
where once again $c(v)$ denotes the coordination number of the vertex $v$, equal to the number of half lines of the intermediate-fields hooked to this vertex, and $\epsilon_k=\pm 1$ is a sign depending of the orientation of the momentum. As a result, the bosonic integral \eqref{bosonicint} becomes:
\begin{align}\label{Gaussianformula}
\int d\nu_{\mathcal{B}}\bigg[&\prod_{m\in\mathcal{B}}e^{-V_j}(\lambda)^{c(m)}\sum\limits_{\substack{\{x_l^{(m)}\}\\\sum_{l\geq1}lx_l^{(m)}=c(m)}}\frac{c(m)!}{\prod_{l\geq1}x_l^{(m)}!l^{x_l^{(m)}}}\prod_{l\geq 1}\overline{[\partial_{\sigma}^l(-V_j)]^{m_l}}\bigg]\,,
\end{align}
where we used again the compact notations for the derivatives, and the integrations over momenta are implicit. Note that the bar over the product means that we extracted the factor $(\lambda)^{l\times x_l^{(m)}}$, with $\lambda^2:=-u/2\pi\beta$. Using the constraint: $\sum_m c(m)=2(|\mathcal{B}|-1)$, with $|\mathcal{B}|$ the number of vertices of $\mathcal{B}$, and because of the bound $\Vert R \Vert \leq \cos^{-1}(\varphi/2)$, the equation \eqref{Gaussianformula} satisfies the inequality:
\begin{align}
\nonumber\big|\int d\nu_{\mathcal{B}} &F_{\mathcal{B}}(\vec{\tau})\big|\leq \bigg(\frac{\lambda^2}{\cos^2(\phi/2)}\bigg)^{|\mathcal{B}|-1} \int d\nu_{\mathcal{B}}\bigg[\prod_{m\in\mathcal{B}}e^{-V_{j_m}(\vec{\tau}_m)}\\
&\quad\times \sum\limits_{\substack{\{x_l^{(m)}\} \sum_{l\geq1}lx_l^{(m)}=c(m)}}\frac{c(m)!}{\prod_{l\geq1}x_l^{(m)}!l^{x_l^{(m)}}} \prod_{l\geq 1}\overline{\vert\partial_{\sigma}^l(-V_j)\vert^{x_l(m)}}\,\bigg]_{R=\mathrm{id}}\,,
\end{align}
where in the right hand side we set $R=\mathrm{id}$. Defining:
\begin{equation}
G_{\mathcal{B}}:=\prod_{m\in\mathcal{B}}\sum\limits_{\substack{\{x_l^{(m)}\}\\\sum_{l\geq1}lx_l^{(m)}=c(m)}}\frac{c(m)!}{\prod_{l\geq1}x_l^{(m)}!l^{x_l^{(m)}}}
\prod_{l\geq 1}\overline{\vert\partial_{\sigma}^l(-V_j)\vert^{x_l(m)}}_{R=\mathrm{id}}\,,
\end{equation}
and since the Gaussian measure $d\nu_{\mathcal{B}}$ is positive, we can use the Cauchy--Schwarz inequality to get:
\begin{equation}\label{CSbound}
\int d\nu_{\mathcal{B}} \prod_{m\in\mathcal{B}}e^{-V_{j_m}(\vec{\tau}_m)}G_{\mathcal{B}}\leq \bigg(\int d\nu_{\mathcal{B}}\prod_{m\in\mathcal{B}}\big|e^{-2V_{j_m}(\vec{\tau}_m)}\big|\bigg)^{1/2}\bigg(\int d\nu_{\mathcal{B}} \, |G_{\mathcal{B}}|^2\bigg)^{1/2}\,.
\end{equation}
We called the first term \emph{non-perturbative factor}, and the second term \emph{perturbative factor}, and shall treat separately each of them.

\subsubsection{Final bounds}
\label{sectionboundgauss}

\paragraph{Non-perturbative bound.} We begin with the first term of the bosonic integral, the non-perturbative contribution (following the notations of~\cite{Lahoche:2018:ConstructiveTensorialGroup-2}):
\begin{equation}
B_1:=\int \droit\nu_{\mathcal{B}}\prod_{a\in\mathcal{B}}\bigg|e^{-2V_{j_a}(\vec{\tau}_a)}\bigg|,\label{B1}
\end{equation}
$\droit\nu_{\mathcal{B}}$ being the restriction of the measure $\droit\nu_{\mathcal{J}}$ to the block $\mathcal{B}$. We aim at proving the following lemma:
\begin{lemma}\label{boundb1}
For $\vert u\vert$ small enough, $B_1$ defined in \eqref{B1} satisfies the following uniform bound:
\begin{equation}
B_1\leq e^{\mathcal{O}(1) \, \frac{\vert u\vert N}{2\pi\beta}\cos^{-1}(\varphi/2)}\,.
\end{equation}
\end{lemma}
\textit{Proof.} First of all, note that: $\big|e^{-2V_{j_a}(\vec{\tau}_a)}\big|$ is uniformly bounded by $\leq e^{2|V_{j_a}(\vec{\tau}_a)|}$. Secondly, because of the identity:
\begin{equation}
\ln_2(1-x)=\int _0^1\droit t \, \frac{tx^2}{1-tx},
\end{equation}
and using the bound \eqref{boundresolventmlve}, we get the inequality:
\begin{equation}
|V_j|\leq \frac{1}{\cos(\varphi/2)}\bigg|\Tr (\mathcal{H}_{\leq j})^2-\Tr (\mathcal{H}_{\leq j-1})^2\bigg|\,,
\end{equation}
with the definition $\mathcal{H}_{\leq j}:= \chi_{\leq j} \mathcal{H}\chi_{\leq j}$. The cyclicity of the trace gives the relation:
\begin{equation}
\Tr (\mathcal{H}_{\leq j})^2-\Tr (\mathcal{H}_{\leq j-1})^2=\Tr \mathcal{H}(\chi_{\leq j} \mathcal{H}\chi_{\leq j}-\chi_{\leq j-1} \mathcal{H}\chi_{\leq j-1})\,.
\end{equation}
Moreover, because of the definition $\chi_j:=\chi_{\leq j}-\chi_{\leq j-1}$, the difference may be rewritten as:
\begin{equation}
\chi_{\leq j} \mathcal{H}\chi_{\leq j}-\chi_{\leq j-1} \mathcal{H}\chi_{\leq j-1}=\chi_{ j} \mathcal{H}\chi_{\leq j-1}+\chi_{\leq j-1} \mathcal{H}\chi_{ j}
\end{equation}
and we get, using again the fact that the trace is cyclic:
\begin{equation}
\Tr (\mathcal{H}_{\leq j})^2-\Tr (\mathcal{H}_{\leq j-1})^2=2\Tr \,(\mathcal{H}\chi_{ j} \mathcal{H}\chi_{\leq j-1})\,.
\end{equation}
Because $\chi_{\leq j}^2=\chi_{\leq j}$ and $\chi_j^2=\chi_j$, and defining $C_j:=\chi_j C\chi_j$ and $C_{\leq j}:=\chi_{\leq j} C\chi_{\leq j}$, the trace may be explicitly computed as:
\begin{equation}
\Tr \,(\mathcal{H}\chi_{ j} \mathcal{H}\chi_{\leq j-1})=\frac{-uN}{2\pi\beta} \int \droit \textbf{p}^\prime \,\sigma(-\textbf{p}^\prime) \left(\int \droit \textbf{p} \,C_{\leq j} (\textbf{p})C_j(\textbf{p}-\textbf{p}^\prime)\right) \sigma(\textbf{p}^\prime)\,,
\end{equation}
the factor $N$ arising from the trace trace over internal indices $\Tr \textbf{I}=N$. The covariance for bosonic variables is then modified as $X_\mathcal{B}^{-1}\to X_\mathcal{B}^{-1}-8\cos^{-1}(\varphi/2)\mathbb{I}_\mathcal{B}\otimes \hat{L}$, $\hat{L}$ being the bounded operator of $L_2(S^1,\mathbb{R})$:
\begin{equation}
\hat{L}(\textbf{p}^\prime):=\mathcal{O}(1) \, \frac{\vert u\vert N}{2\pi\beta}\int \droit \textbf{p} \,C_{\leq j} (\textbf{p})C_j(\textbf{p}-\textbf{p}^\prime)\,, \qquad \Vert \hat{L}\Vert \leq \mathcal{O}(1) \, \frac{\vert u\vert N}{2\pi\beta} M^{-2j}\,,
\end{equation}
$\mathcal{O}(1)$ denoting some numerical constant of order $1$. Computing the Gaussian integration gives the determinant:
\begin{equation}
\det \left[\mathrm{id}-\frac{8}{\cos(\varphi/2)}\frac{\vert u\vert N}{2\pi\beta}X_\mathcal{B}\otimes \hat{L}\right]^{-1/2} = e^{-\frac{1}{2} \Tr\ln\left(\mathrm{id}-\frac{8}{\cos(\varphi/2)}\frac{\vert u\vert N}{2\pi\beta}X_\mathcal{B}\otimes \hat{L}\right)}\,.
\end{equation}
Given a positive bounded operator $\hat{A}$, we have the following bound:
\begin{equation}
-\Tr\ln(1-\hat{A})=\sum_{n=1}^\infty \frac{\Tr(\hat{A})^n}{n} \leq \Tr(\hat{A})\times \sum_{n=2}^\infty \frac{\Vert \hat{A}\Vert^n}{n} \leq \Tr(\hat{A})\times \sum_{n=2}^\infty {\Vert \hat{A}\Vert^n}\,.
\end{equation}
Setting $\hat{A}:=\frac{8\vert u\vert N}{2\pi\beta}\cos^{-1}(\varphi/2)\mathbb{I}_\mathcal{B}\otimes \hat{L}$, it is easy to check that:
\begin{equation}
\Tr(\hat{A})\leq \mathcal{O}(1)\frac{8\vert u\vert N}{2\pi\beta} \cos^{-1}(\varphi/2) \vert \mathcal{B} \vert\,,\qquad \Vert \hat{A}\Vert \leq \mathcal{O}(1) \, \frac{8\vert u\vert N}{2\pi\beta}\cos^{-1}(\varphi/2)\,,
\end{equation}
leading to:
\begin{equation}
\vert \Tr\ln(1-\hat{A})\vert \leq \vert \mathcal{B}\vert \sum_{n=1}^\infty \left(\mathcal{O}(1)\frac{8\vert u\vert N}{2\pi\beta}\cos^{-1}(\varphi/2)\right)^n\,.
\end{equation}

\begin{flushright}
$\square$
\end{flushright}

\paragraph{Perturbative bound.} We now move on to the perturbative bound:
\begin{equation}\label{perturbative}
B_2:=\bigg(\int \droit\nu_{\mathcal{B}} \, |G_{\mathcal{B}}|^2\bigg)^{1/2}.
\end{equation}
and we have to prove the lemma:
\begin{lemma}\label{boundb2}
For $\vert\lambda\vert$ small enough, the perturbative contribution $B_2$ satisfies the following bound, up to a global conservation factor:
\begin{equation}
|B_2|\leq \sqrt{(4|\mathcal{B}|-4)!!}\times \prod_{m\in\mathcal{B}}\left(\mathcal{O}(1)\pi\right)^{c(m)}c(m)!M^{-2j_m} \,.
\end{equation}
\end{lemma}
\textit{Proof.} For $l>1$ we have:
\begin{equation}
\int \droit \textbf{p}\,C_{ j}^l(\textbf{p})\leq \int \droit \textbf{p}\,C_{ j}^2(\textbf{p})\leq \frac{1}{M^{4(j-1)}} \int \droit \textbf{p}\,\chi_j(\textbf{p})\,.
\end{equation}
The last integral is nothing but the volume between the disks of radius $\textbf{p}^2=M^{2(j-1)}$ and $\textbf{p}^2=M^{2j}$, as a result:
\begin{equation}
\int \droit \textbf{p}\,C_{ j}^l(\textbf{p}) \leq \mathcal{O}(1) \, \frac{\pi}{M^{2j-4}}\,.
\end{equation}
The result may be generalized for integrals of the type:
\begin{equation}
\mathcal{I}_{\textbf{p}_1,\cdots,\textbf{p}_l}:=\int \droit \textbf{p} \,(\underbrace{C_{\leq j} (\textbf{p}) C_{\leq j} (\textbf{p}-\textbf{p}_1)\cdots C_{\leq j} (\textbf{p}-\textbf{p}_l)}_\textrm{l times}-( {j}\to {j-1})\,.
\end{equation}
The domain of the integral is the intersection of disks centered in $\textbf{0}$, $\textbf{p}_1$, $\cdots$, $\textbf{p}_l$. In the worst case, $\textbf{p}_1=\textbf{p}_2=\cdots=\textbf{0}$, we get:
\begin{equation}
\mathcal{I}_{\textbf{p}_1,\cdots,\textbf{p}_l} \leq M^{-2(l+1)(j-1)}\int \droit \textbf{p} \, \chi_j(\textbf{p})\leq \mathcal{O}(1) \frac{\pi}{M^{-2lj+2}} \leq \mathcal{O}(1) \frac{\pi}{M^{2j-4}}\,.\label{inequalityuseful}
\end{equation}
To compute the Gaussian integral, we have to retain only the leaves of the trees, each of them involving a single intermediate field. In more details, because of the inequality \eqref{inequalityuseful}, the Gaussian block has to be uniformly bounded by a Gaussian integral of the form:
\begin{equation}\label{Gaussianexemple}
H_{\mathcal{B}}:=\int d\bar{\nu}_{\mathcal{B}}\prod_{a\in\mathcal{B}} \left(\mathcal{O}(1) \frac{\pi}{M^{2j_a-4}}\right)^{k_a}\times (\sigma_a)^{k_a}.
\end{equation}
Because we bounded the integrals over each loop with the worst bound, the remaining Gaussian integrations are momentum-independent. Then, the variables $\sigma_a$ do not depend on the momentum, and the Gaussian measure $\int d\bar{\nu}_{\mathcal{B}}$ has covariance $X_{\mathcal{B}}$. Because $||X_{\mathcal{B}}||\leq 1$, it follows that the Gaussian integration \eqref{Gaussianexemple} is bounded by:
\begin{align}
|H_{\mathcal{B}}|\leq \prod_{a\in\mathcal{B}}\left(\mathcal{O}(1) \frac{\pi}{M^{2j_a-4}}\right)^{k_a}\times \bigg(\sum_{a\in\mathcal{B}}k_a\bigg)!! \,.
\end{align}
From equation \eqref{perturbative}, the double factorial is trivially bounded by $(4|\mathcal{B}|-4)!!$. As a result, assuming $j_m > 2$, we get the following bound\footnote{Note that this restriction is unnecessary, the first slices can be bounded with a simple loop vertex expansion.} (up to a global conservation factor):
\begin{align}
\nonumber |B_2|\leq \sqrt{(4|\mathcal{B}|-4)!!}\times \prod_{m\in\mathcal{B}}\sum\limits_{\substack{\{x_l^{(m)}\}\\\sum_{l\geq1}lx_l^{(m)}=c(m)}}&\frac{c(m)!}{\prod_{l\geq1}x_l^{(m)}!l^{x_l^{(m)}}}\left\vert\lambda\right\vert^{\frac{x_1^{(m)}}{2}}\left(\mathcal{O}(1)\pi\right)^{c(m)}M^{-(2j_m-4)}\,.
\end{align}
Assuming $|\lambda|\leq 1$,
\begin{equation}\label{rq}
\sum\limits_{\substack{\{x_l^{(m)}\}\\\sum_{l\geq1}lx_l^{(m)}=c(m)}}\frac{1}{\prod_{l\geq1}x_l^{(m)}!l^{x_l^{(m)}}}
\end{equation}
is nothing but the coefficient of $x^{c(m)}$ in the Taylor expansion of $\prod_{k}e^{x^k/k}=1/(1-x)$.
\begin{flushright}
$\square$
\end{flushright}

As a result, taking into account Lemmas \ref{boundb1} and \ref{boundb2}, we deduce the final bound:
\begin{align}
\nonumber\big|\int d\nu_{\mathcal{B}} \prod_{m\in\mathcal{B}}e^{-V_{j_m}(\vec{\tau}_m)}G_{\mathcal{B}}\big|
&\leq e^{\mathcal{O}(1)\big|\frac{\lambda}{\cos(\phi/2)}\big||\mathcal{B}|}\sqrt{(4|\mathcal{B}|-4)!!}\\
&\times \prod_{m\in\mathcal{B}}c(m)!\left(\mathcal{O}(1)\pi\right)^{c(m)}M^{-2j_m}\,,
\end{align}
and the following proposition for the bosonic integration:
\begin{proposition} \label{boundbosons}
For $\vert\lambda\vert$ small enough, the bosonic integration admits the bound:
\begin{align}
\nonumber\big|\int d\nu_{\mathcal{B}} F_{\mathcal{B}}(\vec{\tau})\big|\leq &\ e^{\mathcal{O}(1)\big|\frac{\lambda}{\cos(\varphi/2)}\big||\mathcal{B}|}\bigg\vert\frac{\lambda}{\cos^2(\varphi/2)}\bigg\vert^{|\mathcal{B}|-1}N^{\vert \mathcal{B}\vert}\\
&\times\sqrt{(4|\mathcal{B}|-4)!!}\times \prod_{m\in\mathcal{B}}c(m)!\left(\mathcal{O}(1)\pi\right)^{c(m)}M^{-2j_m}\,.
\end{align}
\end{proposition}
Finally, collecting the results for bosonic and Grassmann bounds, we get the uniform bound for the free energy $\ln(Z)$ (to simplify the expression we forget the exponential factor $e^{\mathcal{O}(1)\big|\frac{\lambda}{\cos(\varphi/2)}\big||\mathcal{B}|}$):
\begin{equation}
\label{firststep}
\begin{aligned}
|\ln\mathcal{Z}[J,\bar{J},\lambda]|
&\leq \sum_{n=1}^{\infty}\frac{1}{n!} \sum_{\mathcal{J}\,tree}
\bigg[\prod_{k=1}^n\sum_{j_k=0}^{j_{max}}
\bigg]
2^{L(\mathcal{F}_F)}
\bigg(\prod_{\ell_f\in\mathcal{F}_F} \delta_{j_{s(\ell_f)}j_{t(\ell_f)}}\bigg)
\\
&\quad \times
\prod_{\mathcal{B}}
\prod\limits_{\substack{m,m^{\prime}\in\mathcal{B}\\m\neq m^{\prime}}}(1-\delta_{j_{m}j_{m^{\prime}}})N^{\vert \mathcal{B}\vert}\bigg(\frac{|\lambda|}{\cos^2(\varphi/2)}\bigg)^{|\mathcal{B}|-1}
\\
&\quad \times
\sqrt{(4|\mathcal{B}|-4)!!}
\prod_{m\in\mathcal{B}}c(m)!\left(\mathcal{O}(1)\pi\right)^{c(m)}M^{-2j_m}\,,
\end{aligned}
\end{equation}
the factor $2^{L(\mathcal{F}_F)}$ involving the number of fermionic edges $L(\mathcal{F}_F)$. Because of \emph{Cayley's Theorem}, the number of trees with $n$ labelled vertices and coordination numbers $c_i$ for each vertex $i=1,\ldots,n$ is $(n-2)!/ \prod_i(c_i-1)!$, and the sum involved in \eqref{firststep} obeys to
\begin{equation}
\sum\limits_{\substack{c(m) \\ \sum\limits_m c(m) = 2 |\mathcal{B}|-2}} \prod_{m\in\mathcal{B}}c(m)=\frac{(3|\mathcal{B}|-3)!}{(|\mathcal{B}|-2)!(2|\mathcal{B}|-1)!}\; .
\end{equation}
Collecting all the factorials and using Stirling's formula, we get:
\begin{equation}\label{boundcomb}
2\sqrt{(4|\mathcal{B}|-4)!!} \, \frac{(3|\mathcal{B}|-3)!}{(2|\mathcal{B}|-1)!}\leq (|\mathcal{B}|-1)! \, 3^{|3\mathcal{B}|}e^{-|\mathcal{B}|}|\mathcal{B}|^{|\mathcal{B}|}\,.
\end{equation}
We now move on to sum over scale attributions, taking into account the hard core constraint. As explained in full details in~\cite{Gurau:2015:MultiscaleLoopVertex}, the hard core constraint imposes that the scale assignments of vertices in a same block are all different, implying that:
\begin{equation}
\sum_{m\in\mathcal{B}}(j_m-2)\geq \frac{1}{2}\sum_{m\in\mathcal{B}}j_m+\frac{j_{min}-2}{2}|\mathcal{B}|+\frac{|\mathcal{B}|(|\mathcal{B}|-5)}{4},
\end{equation}
where we have introduced explicitly the minimal scale $j_{min}>2$. Then:
\begin{align}
\sum_{\{j_m\}}\prod\limits_{\substack{m,m^{\prime}\in\mathcal{B}\\m\neq m^{\prime}}}(1-\delta_{j_{m}j_{m^{\prime}}})\prod_{m\in\mathcal{B}}M^{-j_m}\leq \bigg(\sum_{j=j_{min}}^{j_{max}}M^{-j/2}\bigg)^{|\mathcal{B}|}\frac{1}{M^{\frac{j_{min}-2}{2}|\mathcal{B}|+\frac{|\mathcal{B}|(|\mathcal{B}|-5)}{4}}}
\end{align}
which, for $j_{min}>2$ and $M>4$, is uniformly bounded by $M^{-|\mathcal{B}|^2/4}$. Note that the upper bound $j_{max}$ can now be sent to infinity without any divergence, ensuring that the theory is well-defined in the ultraviolet limit. \\

The final step is to sum over the fermionic forest. Such a forest can be partitioned into components of cardinal $b_k$, associated to connected blocks of size $k$, having $k$ sub-vertices. The number of fermionic edges is then $\sum_kb_k-1$, and for each component with $k$ sub-vertices, there are $n^{b_k}$ ways to hook a fermionic edge. Moreover, Cayley's theorem (without constraint on the coordinate number) states that the number of trees with $v$ labelled vertices is $v^{v-2}$, leading to a contribution $n^{\sum_kb_k-2}$. Finally, because of the constraint $\sum_{k}kb_k=n$, when the number of (sub) vertices is fixed to $n$, and from Stirling formula: $n^{(\sum_kb_k)-2}\leq (\sum_kb_k)!e^n$, we find:
\begin{align}
\nonumber|\ln\mathcal{Z}[J,\bar{J},\lambda]|&\leq N \sum_{n}\frac{1}{n!}\sum\limits_{\substack{\{b_k\}\\\sum_kkb_k=n}}\frac{n!}{\prod_kb_k!(k!)^{b_k}}2^{\sum_kb_k-1}k^{(\sum_kb_k)-2}\prod_kn^{b_k}\\
&\times\prod_k\bigg[\bigg(\mathcal{O}(1)\frac{\pi^2|\lambda| N}{\cos^2(\varphi/2)}\bigg)^{k-1}\sqrt{(4k-4)!!}\frac{(3k-3)!}{(2k-1)!}M^{-k^2/4}\bigg]^{b_k} .
\end{align}
Taking into account the bound \eqref{boundcomb} and performing the sum over the $\{b_i\}$, we get finally:
\begin{align}\label{finalstep}
|\ln\mathcal{Z}[J,\bar{J},\lambda]|&\leq \sum_{b\geq 0}\bigg[\sum_{n\geq1}\bigg(\frac{\pi^2|\lambda| N}{\cos^2(\varphi/2)}\bigg)^{n-1}3^{3n}n^nM^{-n^2/4}\bigg]^b.
\end{align}
The power of $M$ ensures that, for $M$ sufficiently large, this factor compensates the bad divergence associated to $n^n$, ensuring analyticity in the interior of a cardioid domain $\vert \lambda\vert \leq \frac{\rho}{N} \cos (\varphi/2)$ for small enough $\rho$. The final check to prove Borel summability follows straightforwardly the Section \ref{secBorel}.

\subsection{Fermions in dimensions \texorpdfstring{$d \leq 1$}{d =< 1}}

In this section, we investigate the Borel summability of quartic fermionic models. Note that the partition function of fermions in $d = 0$ at finite $N$ is polynomial (and thus analytic) in the coupling constant since the fermions are Grassmann valued. As a consequence, the partition function $Z$ is analytic (but not polynomial): thus, the main non-trivial content of our result lies in the large $N$ limit. We briefly discuss the case $d=0$, as an instructive step to extend the proof for $d=1$. Moreover, we focus on Majorana fields to anticipate the proof of the constructive bound for the SYK model in the next section, but the extension to Dirac fields is straightforward. To get a well i.i.d real field model, we have to use a physical feature of fermionic fields. Indeed, physically relevant fermionic fields share some internal discrete degrees of freedom, like spin. Assuming that such a spin exists for our fields, we can introduce a new index $\sigma=\pm$ such that $\psi_i$ becomes a bi-vector: $\psi_i^T\equiv (\psi_{i,+},\psi_{i,-})$; and we will denote the components as $\psi_{i,\sigma}$. Thus, the total number of degrees of freedom equals $2N$ in this simplest model; with the following kinetic action:
\begin{equation}
S_{\text{kin}}(\Psi):=\sum_{i,\sigma,j,\sigma^\prime} \psi_{i,\sigma} \left( K_{ij}^{-1} \tau_{\sigma,\sigma^\prime}\right) \psi_{j,\sigma^\prime}\,,
\end{equation}
where the matrix $\tau=\tau^{-1}$ couples the up and down spins, and we must set $\tau=\sigma_1$, the first Pauli matrix. In zero dimension, real fields correspond to Majorana spinors, and $\tau$ may be interpreted as a charge conjugation matrix. We can then use the same strategy as for the bosonic case. We can diagonalize the matrix $\mathcal{W}_{IJ}$ for the blocks indices $I=(ij)$ and $J=(kl)$, denoting the eigenvalues as $\xi_L^2$ :
\begin{equation}
\mathcal{J}_{IJ}=\sum_L\xi_L^2 O_{IL}O_{LJ}^T\,.
\end{equation}
For $d=0$ and $K ^{-1}_{ij}=\delta_{ij}$, the strategy used for the bosonic bound holds. Introducing several intermediate fields, the partition function becomes:
\begin{equation}
Z(u)=\int \droit\mu_C(\psi)\droit\nu(\sigma) \prod_{L}e^{\sqrt{\frac{{-u}}{12} }\xi_L \Xi_L\sigma^{(L)}}\,,
\end{equation}
where $\Xi_L:=\sum_{I=ij}O_{IL}\psi_i\psi_j$ and $d\mu_C(\psi)$, $\droit\nu(\sigma)$ are normalized Gaussian measures. Using the standard result on Berezin integrals:
\begin{equation}
\int \prod_i^{2n} \droit\psi_i \exp\left(\Psi^TA\Psi\right)= 2^n \sqrt{\det A}\,,
\end{equation}
we get the following effective bosonic theory:
\begin{equation}
Z(u,J)=\int \droit\nu(\chi) e^{\mathcal{V}(\chi)}.
\end{equation}
Except for the difference of sign in front of the effective interaction $\mathcal{V}$, it is exactly what we obtained for the bosonic fields. In particular, our conclusions on the bound of absolute values for terms involved in the tree expansion of the free energy hold. The only difference comes from the presence of the matrix $\tau$. From Lemma \ref{lemmauseful2}, it follows that the norm of the resolvent $R(\chi)$ remains bounded as $\cos^{-1}(\varphi/2)$. However, the trace over external indices, "$\Tr$", involved in the definition of the vertex function $\mathcal{V}(\chi)$ becomes a trace over the whole index structure, including spin indices. Each derivative involved in the definition of the effective vertex introduces a $\tau$ insertion. Let $v$ and $w$ be two effective vertices and $\ell$ a link between them. Denoting by $\tr$ the trace over spin indices, the trace $\tr [R^{(v)}\tau R^{(w)} \tau]$ corresponds to the edge $\ell$. The external index structure with respect to the indices $i$, $j$ remaining unchanged, the bound given for bosons still holds. Fixing the external indices, and because the operator norm of $\tau$ is $\Vert \tau \Vert =1$, the traces over spin indices can be bounded as:
\begin{equation}
\vert \tr [R^{(v)}\tau R^{(w)} \tau]\vert \leq \tr \vert R^{(v)}\, R^{(w)}\vert \leq \tr \vert R^{(v)} \vert \times \tr\vert R^{(w)}\vert \,.
\end{equation}
As a consequence, the trace $\Tr \vert R^{c(v)}\vert$ involved in the bosonic bounds becomes $\Tr ( \tr \vert R\vert)^{c(v)}\leq \Vert R\Vert \tr\textbf{I}_2\times \Tr \mathbf{I}= \Vert R\Vert\times 2N$; where we introduced a subscript $2$ for the $2\times 2$ identity matrix $\textbf{I}_2$. Taking into account this new factor $2$ coming from traces over internal indices, this implies the existence of a non-vanishing analyticity domain:
\begin{equation}
\vert u\vert \leq \frac{3}{N\, \xi_0^2}\sqrt{\frac{e}{\pi}} \cos^2 \left(\frac{\varphi}{2}\right)\,,
\end{equation}
a conclusion which can be easily extended to Dirac fermions, or for an internal space with dimension larger than $2$.

\medskip

Let us move to the $d=1$ case. The Fourier representation of the propagator  $C(\omega)$ of the theory is:
\begin{equation}
C(\omega)=\frac{1}{i\omega}\,.
\end{equation}
Note that the model does not require a mass term to be UV regularized because of the boundary condition $\psi_i=-\psi_i(t+\beta)$, which excludes the zero-frequency mode $\omega=0$. The spectrum for fermions is given by formula \eqref{frequencies}:
\begin{equation}
\label{eq:sum-1/omega}
\sum_\omega \frac{1}{i\omega}=\frac{\beta}{2i\pi}\sum_n \frac{1}{n+1/2} =\frac{\beta}{2}\,.
\end{equation}

\noindent
As for bosons, bounding the amplitude requires to bound sums like:
\begin{equation}
f(\omega):=\sum_{\omega^\prime,\omega^{\prime\prime},i} \frac{1}{\omega^{\prime\prime}} \delta(\omega-\omega^{\prime\prime}+\omega^{\prime}) \mathcal{V}_{ii}[\chi_1](\omega^{\prime},\omega^{\prime\prime}):=\Tr B(\omega) {R}[\chi_1]\,,
\end{equation}
with
\begin{equation}
B_{\omega^\prime\omega^{\prime\prime}}(\omega):=\frac{1}{\omega^{\prime\prime}} \delta(\omega-\omega^{\prime\prime}+\omega^{\prime}) \,.
\end{equation}
There is an additional subtlety coming from the fact that the propagator is not real. Thus, redefining the resolvent as:
\begin{equation*}
R^{-1}_{\omega\omega^{\prime}} =\delta_{\omega,\omega^\prime} - \sqrt{\frac{u}{3\beta}}\, C^{1/2}(\omega)\Sigma_{\omega\omega^{\prime}}C^{1/2}(\omega^\prime)=\delta_{\omega,\omega^\prime}+i\sqrt{\frac{u}{3\beta}}\frac{1}{\sqrt{\omega}}\Sigma_{\omega\omega^{\prime}}\frac{1}{\sqrt{\omega^\prime}}\,.
\end{equation*}
The operator
\begin{equation}
\label{eq:1d-fermion-H}
\mathcal{H}:=\frac{1}{\sqrt{\omega}}\Sigma_{\omega\omega^{\prime}}\frac{1}{\sqrt{\omega^\prime}}\,,
\end{equation}
is Hermitian due to the reality condition $\Sigma_{\omega\omega^{\prime}}^*=\Sigma_{-\omega,-\omega^{\prime}}$. As a result, $R$ admits the operator bound:
\begin{equation}
\Vert R\Vert \leq \cos^{-1}\left(\frac{\varphi}{2}\right)\,,
\end{equation}
with $\varphi=\arg(u)$. Note that, due to the sign in front of $u$ in the definition \eqref{sykdef}, the cardioid domain is rotated by $\pi$. Because $B(\omega)$ is not positive defined, we do not use the same trick as in the previous section to bound $f(\omega)$. However, we can use the properties of the operator bound. In particular, let $U$ be the unitary transformation acting on the Hilbert space $\mathbb{C}^N\otimes L_2(S_1,\mathbb{R})$ which diagonalizes $\mathcal{V}_I$, and let $r_{i\omega}$ be its eigenvalues. $f(\omega)$ may be rewritten as:
\begin{equation}
f(\omega)=\sum_{\omega^\prime,i} r_{i\omega^\prime} \tilde{B}_{i;\omega^\prime}(\omega)
\end{equation}
where $\tilde{B}_{i;\omega^\prime}(\omega):= (U^\dagger B(\omega) U)_{i,i; \omega^\prime\omega^\prime}$. Obviously, $\tilde{B}_{i;\omega^\prime} \in \mathbb{C}^N\otimes L_2(S_1,\mathbb{R})$. Indeed:
\begin{equation}
\Vert \tilde{B}(\omega) \Vert_2=\sum_{i,\omega^\prime} \tilde{B}_{i;\omega^\prime}^\dagger \tilde{B}_{i;\omega^\prime} \leq \left(\sum_{i,\omega^\prime} \vert \tilde{B}_{i;\omega^\prime}\vert \right)^2 \leq \left(\frac{\beta}{2}\right)^2\,.
\end{equation}
As a result, because $R$ is bounded and from definition of the operator norm, we get:
\begin{equation}
\vert f(\omega) \vert \leq \Vert R\Vert \times \Vert \tilde{B}(\omega) \Vert_2 = \frac{\beta}{2}\,\cos^{-1}\left(\frac{\varphi}{2}\right)\,.
\end{equation}
Then, the bound deduced in the previous section for $1d$ bosons holds, up to the replacements $\beta \coth (\beta m/2)/2m\to \beta/2$ and $\sqrt{u/2\beta} \to \sqrt{u/3\beta}$; and taking into account the $1/2$ per vertex coming from the substitution $e^{-\Tr\ln}\to e^{\frac{1}{2}\Tr\ln}$. We thus proved the existence of a finite analytic domain with shape of a cardioid.

\subsection{Disordered Majorana fermions in \texorpdfstring{$d = 1$}{d = 1}: the SYK model} \label{sec:exp-1d:fermion}

We proved the existence of a finite analyticity domain for the fermionic part of the SYK model with a fixed coupling tensor. In this section, we will complete the proof of the Proposition \ref{Stat2}, taking into account the disordered nature of the coupling tensor $\mathcal{J}_{ijkl}$. In particular, we show that the Gaussian integration preserves the existence of a uniform analyticity domain in the large $N$ limit.

\begin{figure}[ht]
\begin{center}
\includegraphics[scale=1]{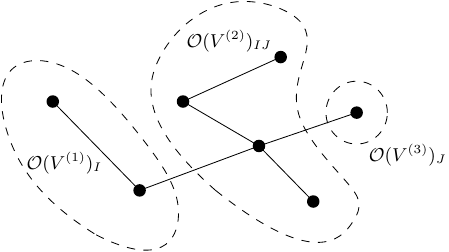}
\caption{A typical tree contributing to the constructive expansion of the fermionic integral.}
\label{figtreecompo}
\end{center}
\end{figure}

\medskip
Let us return on the building of the first bound, see equation \eqref{amplitudeutile}. Let $\mathcal{T}$ be the tree pictured on the Figure \ref{figtreecompo}, with $n$ vertices, and let $\mathcal{O}(V^{(1)})_I$, $\mathcal{O}(V^{2})_{IJ}$ and $\mathcal{O}(V^{(3)})_J$ three connected components such that the amplitude $\mathcal{A}_{\mathcal{T}}$ reads (we are only interested by the internal index structure of the amplitude, and we forget the sums over frequencies $\omega$):
\begin{equation}
\mathcal{A}_{\mathcal{T}} \propto \sum_{IJ} \mathcal{O}(V^{(1)})_I \,\xi^2_I\, \mathcal{O}(V^{2})_{IJ} \,\xi^2_J\, \mathcal{O}(V^{(3)})_J\,.
\end{equation}

\begin{figure}[ht]
\begin{center}
$\underset{a}{\vcenter{\hbox{\includegraphics[scale=0.75]{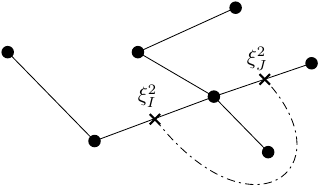} }}}\quad \underset{b}{\vcenter{\hbox{\includegraphics[scale=0.75]{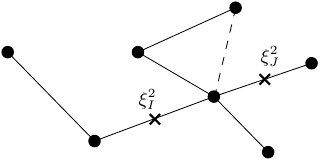} }}}\quad \underset{c}{\vcenter{\hbox{\includegraphics[scale=0.75]{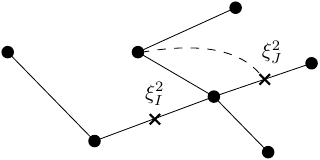} }}}$
\caption{Allowed Wick-contractions on the tree $\mathcal{T}$. Contraction between two edge variables (a), between two corners (b) and between edge and corner (c). }
\label{figAllow}
\end{center}
\end{figure}

Due to Wick theorem, the computation of the Gaussian integral over $\mathcal{J}$ generates "pairings" between $\xi^2_I$ variables, and then adds loops on the tree. There are two sources for $\xi^2_I$ variables: the resolvent and the links, generating three types of pairing:
\begin{enumerate}
\item Wick contractions between two edge variables represented as the dashed-dotted line on Figure \ref{figAllow} (a).

\item Wick contractions between two resolvents, as on Figure \ref{figAllow} (b).

\item Wick contractions between resolvent and edge variables, as on Figure \ref{figAllow} (c).
\end{enumerate}

Wick contractions of type (2) and (3) are effective, in the sense that the resolvent is a function of $\xi^2_I$. To clarify the graphical notations, we will picture the ordinary Wick contractions with dashed-dotted lines and the effective ones with dashed lines on Figure \ref{figAllow}. Ultimately, we are interested to deal with the limit $N\to \infty$, and some of the allowed contractions will be discarded in this limit. For instance, the contractions of type (1) between edge variables introduce additional Kronecker delta, which reduces the ranges of the sums over internal indices, and ultimately the final dependence on $N$ of the amplitude $\mathcal{A}_{\mathcal{T}}$. The same thing holds for contractions between edge variable and resolvent which are not hooked on the boundary of the corresponding edge, as well as between different resolvents.

\begin{figure}
\begin{center}
\includegraphics[scale=0.6]{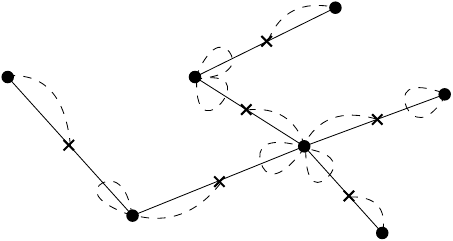}
\caption{A typical tree contributing to the averaged amplitude $\langle \mathcal{A}_{\mathcal{T}}\rangle_{\mathcal{J}}$. }\label{figtreecont}
\end{center}
\end{figure}

The contractions of the edge variables which optimize the $N$ dependence of $\mathcal{A}_{\mathcal{T}}$ are then necessarily between one of the two resolvents hooked on the boundary of the corresponding edge; and the remaining contractions have to be "self loops" between resolvents. As a result, for each edge, there are two allowed contractions:\\
\begin{equation}
\vcenter{\hbox{\includegraphics[scale=1.1]{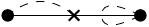}}}\quad + \quad \vcenter{\hbox{\includegraphics[scale=1.1]{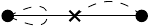}}}\,, \label{rightleft}
\end{equation}
a typical tree contributing to $\langle \mathcal{A}_{\mathcal{T}}\rangle_{\mathcal{J}}$ -- the averaged amplitude is pictured on Figure \ref{figtreecont}.\footnote{The notation $\langle X \rangle_{\mathcal{J}}$ means Gaussian averaging with respect to the tensor coupling $\mathcal{J}$, normalized to $1$: $\langle 1 \rangle_{\mathcal{J}}=1$}
$\langle \mathcal{A}_{\mathcal{T}}\rangle_{\mathcal{J}}$ is such a sum over all permutations corresponding to allowed contractions describing in equation \eqref{rightleft}. There are two type of contractions per edges, and $n-1$ edges. We denote as $\langle \mathcal{T}\rangle_{\mathcal{J}}$ the corresponding $2^{n-1}$ graphs obtained from $\mathcal{T}$. Therefore, $\langle \mathcal{A}_{\mathcal{T}}\rangle_{\mathcal{J}}$ is bounded by $2^{n-1}\times \sup\{ \vert \mathcal{A}_G \vert, G\in \langle \mathcal{T}\rangle_{\mathcal{J}} \}$, where $ \mathcal{A}_G$ denotes the amplitude for the graph $G$. The final step to bound $\langle \mathcal{A}_{\mathcal{T}}\rangle_{\mathcal{J}}$ is then to bound the amplitudes $\mathcal{A}_G$, and we have the following statement:
\begin{lemma}
The averaged amplitudes $\mathcal{A}_{G}$, for $G\in \langle \mathcal{T}\rangle_{\mathcal{J}}$ must be of order $\sqrt{N}$:
\begin{equation}
\left\vert \langle \mathcal{A}_{\mathcal{T}_n}\rangle_{\mathcal{J}} \right\vert = \mathcal{O}(\sqrt{N})\,.
\end{equation}
\end{lemma}

\noindent
\emph{Proof.} To prove this, we proceed recursively on $n$. Let us consider the simpler tree $\mathcal{T}_2$ with two vertices, on the right of equation \eqref{rightleft}. Writing $\tilde R_{ij}= R_{ij}\delta_{ij}$, the amplitude becomes proportional to:
\begin{equation}
\left\vert \langle \mathcal{A}_{\mathcal{T}_2}\rangle_{\mathcal{J}} \right\vert \propto\left\vert \sum_{I}\langle \tilde{R}_I \rangle_{\mathcal{J}} \langle \xi^2_I \tilde{R}_I \rangle_{\mathcal{J}} \right\vert \leq \sqrt{\sum_I \langle \vert \tilde{R}_I\vert \rangle_{\mathcal{J}}^2 \times \sum_I \langle \vert \xi^2_I \tilde{R}_I \vert\rangle_{\mathcal{J}}^2}\,.
\end{equation}
Due to the normalization of the Gaussian integration, the first factor $\sum_I \langle \vert \tilde{R}_I\vert \rangle_{\mathcal{J}}^2$ is uniformly bounded as $\cos^{-2}\left( \frac{\varphi}{2}\right) \Tr\,\textbf{I}$. For the second factor, the Cauchy-Schwarz inequality leads to:
\begin{equation}
\sum_I \langle \vert \xi^2_I \tilde{R}_I \vert\rangle_{\mathcal{J}}^2 \leq {\sum_I \langle(\xi^2_I)^2\rangle_{\mathcal{J}} } {\sum_I \langle\vert \tilde{R}_I\vert^2\rangle_{\mathcal{J}} } \propto \frac{\vert \bar{g}\vert^2}{\cos^2\left( \frac{\varphi}{2}\right) } {N^{-3}\,N^2}\times {N}\,. \label{recur1}
\end{equation}
As a result $\left\vert \langle \mathcal{A}_{\mathcal{T}_2}\rangle_{\mathcal{J}} \right\vert = \mathcal{O}(\sqrt{N})$.

\medskip
Now, assuming that the property holds for any tree of size $n$, we have to prove that it survives after adding a leaf. The amplitude $\mathcal{A}_{\mathcal{T}_{n+1}}$ reads:
\begin{equation}
\mathcal{A}_{\mathcal{T}_{n+1}}=\sum_I V_I^{(n)} \xi^2_I \tilde{R}_I\,, \label{sumzero}
\end{equation}
where $V_I^{(n)}$ denotes the rest of the tree with $n$ vertices. Denoting as $v$ the vertex to which the leaf is added and as $c(v)+1$ its coordination number, $V_I^{(n)}$ has the following structure:
\begin{equation}
V_I^{(n)}= \sum_{\{I_k\}} V_{I; I_1,\cdots I_{c(v)}} \prod_{k=1}^{c(v)} \xi_{I_k}^2V_{I_k}^{(m(k))} \,.
\end{equation}
Along each of the $c(v)$ edges, we can bound the expression using Cauchy-Schwarz inequality. For $k=1$, we define
\begin{equation}
V_{I_1}^{(n)}= \sum_{\{I_k\},k\neq 1,I} V_{I; I_1,\cdots I_{c(v)}} \prod_{k=2}^{c(v)} \xi_{I_k}^2V_{I_k}^{(m(k))}\xi^2_I \tilde{R}_I\,.
\end{equation}
Then, assuming that we averaged on $\mathcal{J}$ only on the connected component $V_{I_1}^{(m(1))}$, we have two configurations, corresponding to the two allowed Wick contractions along the edge $1$. However, even in the case when the contraction is on the component $V_{I_1}^{(n)}$, we can replace the contracted resolvent with the resolvent of the component $V_{I_1}^{(m(1))}$, so that in any case we have the bound:
\begin{align}
\nonumber \sum_{I_1} \langle V_{I_1}^{(n)} \rangle_{\mathcal{J}} \langle \xi_{I_1}^2V_{I_1}^{(m(1))} \rangle_{\mathcal{J}}& \leq \sum_{I_1} \langle \vert V_{I_1}^{(n)} \vert \rangle_{\mathcal{J}} \sqrt{ \sum_{I_1} \langle \vert \xi_{I_1}^2V_{I_1}^{(m(1))}\vert \rangle_{\mathcal{J}}^2 }\,.
\end{align}
Recursively, we get:
\begin{equation}
\sum_{I_1} \langle V_{I_1}^{(n)} \rangle_{\mathcal{J}} \langle \xi_{I_1}^2V_{I_1}^{(m(1))} \rangle_{\mathcal{J}}\leq \sum_{\{I_k\}} \langle \vert V_{I; I_1,\cdots I_{c(v)}} \vert \rangle_{\mathcal{J}} \prod_{k=1}^{c(v)} \mathcal{C}_k\,,
\end{equation}
with: $\mathcal{C}_k:=\sqrt{ \sum_{I_1} \langle \vert \xi_{I_k}^2V_{I_k}^{(m(k))}\vert \rangle_{\mathcal{J}}^2 }$. Finally, returning on the sum \eqref{sumzero}, we get:
\begin{equation}
\sum_I \langle \vert V_I^{(n)} \xi^2_I \tilde{R}_I \vert \rangle_{\mathcal{J}} \leq \prod_{k=1}^{c(v)} \mathcal{C}_k \times \sqrt{\sum_{I,\{I_k\}} \langle \vert V_{I; I_1,\cdots I_{c(v)}} \vert \rangle_{\mathcal{J}} \sum_I \langle \vert \xi^2_I \tilde{R}_I \vert\rangle_{\mathcal{J}}^2}\,.
\end{equation}
From equation \eqref{recur1}, the second term into the square-root is of order $1$. Moreover, the first term is nothing but a big trace, like for \eqref{equationvertex}, as a result:
\begin{equation}
\sum_{I,\{I_k\}} \langle \vert V_{I; I_1,\cdots I_{c(v)}} \vert \rangle_{\mathcal{J}} \lesssim \left(\frac{1}{2} \cos^{-c(v)-1}\left(\frac{\varphi}{2}\right)\right) \times N\,.
\end{equation}
Finally, because of our recursion hypothesis, all the $\mathcal{C}_k$ have to be of order $1$ in $N$, implying $\vert \langle \mathcal{A}_{\mathcal{T}_{n+1}} \rangle_{\mathcal{J}} \vert \sim \sqrt{N}$.

\begin{flushright}
$\square$
\end{flushright}

Therefore, the only changes from the bound for the fermionic part coming from the Gaussian integration are 1) The replacement $N \xi_0^2\to \bar g$ and 2) the combinatorial factor $2^{n-1}$ coming from the Wick contractions. As a result, the averaged amplitude $\langle \mathcal{A}_{\mathcal{T}_n} \rangle_{\mathcal{J}}$ admits the bound:
\begin{equation}
\left\vert \langle \mathcal{A}_{\mathcal{T}_n}\rangle_{\mathcal{J}} \right\vert \leq \bar{K}\sqrt{N}\times \vert {u}\vert^n (\bar{\rho}\,)^n \cos^{-2n} \left(\frac{\varphi}{2}\right) \,,
\end{equation}
meaning that the expansion in analytic in the interior of the cardioid domain $\rho \leq \bar{\rho} \cos^2(\varphi/2)$ for some constant $\bar{K}$, with $(\bar{\rho}\,)^{-1} := 24\sqrt{e/\pi}/ \beta \bar{g}$. Once again, the final check to prove Borel summability follows the Section \ref{secBorel}.

\section*{Acknowledgements}

We are grateful to Vincent Rivasseau for discussions and comments on the last version of the draft.
V.L.\ is also indebted to Vincent Rivasseau for having introduced him to constructive field theory -- and, in particular, to the forest formula -- and for his continuous support.

The work of H.E.\ is conducted under a Carl Friedrich von Siemens Research Fellowship of the Alexander von Humboldt Foundation for postdoctoral researchers.

\section*{Data Availability}

Data sharing is not applicable to this article as no new data were created or analyzed in this study.

\appendix

\section{The Brydges-Kennedy-Abdes\-selam-Rivasseau forest formula} \label{appB}

A forest formula expands a quantity defined on $n$ points in terms of forests built on these points. There are many forest formulas, but the BKAR formula seems to be the only one which is both \emph{symmetric} under permutation of the $n$ points and \emph{positive}. In this appendix, we only recall the theorem without proof, much additional material may be found in~\cite{Abdesselam:1995:TreesForestsJungles, Gurau:2015:MultiscaleLoopVertex}.\\

\noindent
Let $[1, \ldots, n]$ be the finite set of points considered above. An edge $l$ between two elements $i,j\in [1, \ldots, n]$ is a pair $(i,j)$ for $1\leq i<j\leq n$, and the set of such edges can be identified with the set of lines of $K_n$, the complete graph with $n$ vertices. Consider the vector space $S_n$ of $n\times n$ symmetric matrices, whose dimension is $n(n+1)/2$ and the \emph{compact and convex} subset $PS_n$ of \emph{positive} symmetric matrices whose diagonal coefficients are all equal to $1$, and off-diagonal elements are between $0$ and $1$. Any $X\in PS_n$ can be parametrized by $n(n-1)/2$ elements $X_l$, where $l$ run over the edges of the complete graph $K_n$. Let us consider a smooth function $f$ defined in the interior of $PS_n$ with continuous extensions to $PS_n$ itself. The BKAR forest formula states that:
\begin{theorem}[The BKAR forest formula]
\label{BKAR}
We have
\begin{equation}
f(\mathbf{1})=\sum_{\mathcal{F}}\int \droit\mathit{w}_{\mathcal{F}}\partial_{\mathcal{F}}f[X^{\mathcal{F}}(\mathit{w}_{\mathcal{F}})]
\end{equation}
where $\mathbf{1}$ is the matrix with all entries equal to $1$, and:
\begin{itemize}
\item The sum is over the forests $\mathcal{F}$ over $n$ labelled vertices, including the empty forest.

\item The integration over $d\mathit{w}_{\mathcal{F}}$ means integration from $0$ to $1$ over one parameter for each edge of the forest. Note that there is no integration for the empty forest since by convention an empty product is $1$.

\item $\partial_{\mathcal{F}}:=\prod_{l\in\mathcal{F}}\partial_l$ means a product of partial derivatives with respect to the variables $X_l$ associated to the edge $l$ of $\mathcal{F}$.

\item The matrix $X^{\mathcal{F}}(\mathit{w}_{\mathcal{F}})\in PS_n$ is such that $X^{\mathcal{F}}_{ii}(\mathit{w}_{\mathcal{F}})=1\,\forall i$, and for $i\neq j$ $X^{\mathcal{F}}_{ij}(\mathit{w}_{\mathcal{F}})$ is the infimum of the $\mathit{w}_l$ variables for $l$ in the unique path from $i$ to $j$ in $\mathcal{F}$. If no such path exists, by definition $X^{\mathcal{F}}_{ij}(\mathit{w}_{\mathcal{F}})=0$.
\end{itemize}
\end{theorem}

\section{Nevanlinna's theorem}\label{appC}

\begin{theorem}
\label{thm:Nevanlinna}
A series $\sum_{n}\frac{a_n}{n!}\lambda^n$ is \emph{Borel summable} to a function $f(\lambda)$ if the following conditions are met:
\begin{itemize}
\item $f(\lambda)$ is analytic in a disk $Re(\lambda^{-1})>R^{-1}$ with $R\in\mathbb{R}^+$.

\item $f(\lambda)$ admits a Taylor expansion at the origin:
\begin{equation}
f(\lambda)=\sum_{k=0}^{r-1}a_{k}\lambda^k+R_rf(\lambda), \qquad |R_rf(\lambda)|\leq K\sigma^rr!|\lambda|^r, \label{borelbound}
\end{equation}
for some constants $K$ and $\sigma$ independent of $r$.
\end{itemize}
If $f(\lambda)$ is Borel summable in $\lambda$, then:
\begin{equation}
\mathcal{B}=\sum_{n=0}^{\infty}\frac{1}{n!}a_{n}t^n
\end{equation}
is an analytic function for $|t|<\sigma^{-1}$ which admits an analytic continuation in the strip $\{z \mid |Im(z)|<\sigma^{-1}\}$ such that $|\mathcal{B}|\leq Be^{t/R}$ for some constant $B$ and $f(\lambda)$ is represented by the absolutely convergent integral:
\begin{equation}
f(\lambda)=\frac{1}{\lambda}\int_{0}^{+\infty}\droit t \, \mathcal{B}e^{-t/\lambda}.
\end{equation}
\end{theorem}\label{Borel}

\bibliography{constructive_quartic_models_part1.bib}

\end{document}